# Mesh-Agnostic Decoders for Supercritical Airfoil Prediction and Inverse Design

Runze Li,[1] Yufei Zhang,[2] Haixin Chen[3]

(*Tsinghua University, Beijing, 100084, People's Republic of China*)

**Mesh-agnostic models have advantages in terms of processing unstructured spatial data and incorporating partial differential equations. Recently, they have been widely studied for constructing physics-informed neural networks, but they need to be trained on a case-by-case basis and require large training times. On the other hand, fast prediction and design tools are desired for aerodynamic shape designs, and data-driven mesh-based models have achieved great performance. Therefore, this paper proposes a data-driven mesh-agnostic decoder that combines the fast prediction ability of data-driven models and the flexibility of mesh-agnostic models. The model is denoted by an implicit decoder, which consists of two subnetworks, i.e., ShapeNet and HyperNet. ShapeNet is based on implicit neural representation, and HyperNet is a simple neural network. The implicit decoder is trained for the fast prediction of supercritical airfoils. Different activation functions are compared, and a spatial constraint is proposed to improve the interpretability and generalization ability of the model. Then, the implicit decoder is used together with a mesh-based encoder to build a generative model, which is used for the inverse design of supercritical airfoils with specified physical features.**

**Nomenclature**

| | |
|---|---|
| *AoA* | = angle of attack (degree) |
| $\alpha_{\text{TE}}$ | = railing edge slope angle (degree) |
| **b** | = output bias vector (ShapeNet outputs) |
| $\bar{\mathbf{c}}$ | = coefficient vector (HyperNet inputs) |
| **c** | = physical codes (PIVAE latent variables) |
| $\mathbf{c}_0$ | = sample labels |
| $C_L$ | = lift coefficient |
| $C_m$ | = moment coefficient |
| $C_p$ | = pressure coefficient |

---

[1]Postdoctoral research assistant, School of Aerospace Engineering, email: lirz16@tsinghua.org.cn

[2]Associate professor, School of Aerospace Engineering, senior member AIAA, email: zhangyufei@tsinghua.edu.cn

[3]Professor, School of Aerospace Engineering, associate fellow AIAA, email: chenhaixin@tsinghua.edu.cn (Corresponding Author)

| | | |
|---|---|---|
| $\Phi$ | = | implicitly defined function |
| **I** | = | identity matrix |
| $M_\text{w}$ | = | wall Mach number |
| $M_\text{w,1}$ | = | wall Mach number in front of a shock wave |
| $M_\infty$ | = | free-stream Mach number |
| $n_b$ | = | number of spatial bases |
| $n_{\bar{c}}$ | = | number of coefficients |
| $n_c$ | = | dimensionality of physical codes |
| $n_\text{d}$ | = | number of data points |
| $n_\text{l}$ | = | number of latent variables |
| $n_\text{p}$ | = | number of principal components |
| $N_\text{s}$ | = | number of samples |
| $n_t$ | = | number of data points in the template |
| $n_v$ | = | dimensionality of data features |
| $n_x$ | = | dimensionality of the spatial coordinates |
| $n_y$ | = | dimensionality of the model outputs |
| **μ** | = | mean vector |
| $p$ | = | probability density |
| **p** | = | principal component vector |
| $r_\text{LE}$ | = | leading edge radius |
| $Re$ | = | Reynolds number |
| **σ** | = | standard variance vector |
| $t_\text{max}$ | = | maximum relative thickness |
| **θ** | = | model parameters |
| **v** | = | data feature vector (PIVAE latent variables) |
| **W** | = | weights of spatial bases (HyperNet outputs) |
| **x** | = | spatial coordinate vector (ShapeNet inputs) |
| $x$ | = | $x$ coordinate |
| **X** | = | template mesh |
| $X_1$ | = | shock wave location |
| $y$ | = | $y$ coordinate |
| **y** | = | model output vector |
| **Y** | = | sample snapshot |
| $z$ | = | latent variables |
| **Z** | = | spatial bases (ShapeNet outputs) |

## I. Introduction

Machine learning (ML) models have been intensely investigated in the field of aerodynamic shape optimization. ML has been studied for the prediction of aircraft performance or the flow fields around aircraft. It has also been studied for the optimization or inverse design of aerodynamic shapes. These studies and other applications, such as geometric parameterization and dimensionality reduction, were summarized in two reviews [1,2]. Most of the related studies utilized data-driven models, which discover knowledge from data and save time during applications. The recently developed methods have demonstrated great power in two-dimensional airfoil prediction and design cases. However, several challenges still await solutions, i.e., 1) interpretability; 2) generalization ability; 3) the incorporation of flow physics; 4) three-dimensional applications; and 5) efficient sampling. [1-3].

Interpretability usually means making sense of the obtained model or results [4]. Without interpretability, no model performance and robustness guarantees are provided, and the results are often untrustworthy to designers [5]. The generalization ability of a model measures its performance on unseen samples or even unseen domains. Generalization ability and efficient sampling are important because of the scarcity of samples for industrial problems, and big data are usually unavailable for data-driven models [6]. In summary, both interpretability and generalization ability are crucial to improving the credibility of models and their results. Fortunately, both of these characteristics can be improved by incorporating flow physics into the constructed model. For example, interpretability can be improved by incorporating physical features [7], e.g., flow field features, or interpretable bases [8], e.g., principal component analysis (PCA) modes, into the model. A weakly imposed loss function term consisting of the residuals of Navier–Stokes (NS) equations can help minimize the prediction errors of flow fields [9]. Considering these challenges, a promising approach is to use mesh-agnostic models.

Mesh-agnostic models started to attract attention in approximately 2020 for both computer vision and solving partial differential equations (PDEs); the models in these areas are represented by implicit neural representation (INR) [10,11] and the physics-informed neural network (PINN) [12,13], respectively. Consider a class of functions $\Phi$ that satisfy Eq. (1):

$$F(\mathbf{x}, \Phi, \nabla_\mathbf{x}\Phi, \nabla_\mathbf{x}^2\Phi, \cdots) = 0, \ \Phi: \mathbf{x} \mapsto \Phi(\mathbf{x}), \tag{1}$$

where $\mathbf{x} \in \mathrm{R}^{n_x}$ denotes spatial or spatial-temporal coordinates, $F$ is the governing equation, and the derivatives of $\Phi$ with respect to coordinate $\mathbf{x}$ are optional in $F$. Then, $\Phi$ is implicitly defined by the relation defined by $F$. The neural network modeling $\Phi$ is called the INR model, which is a typical mesh-agnostic model. It takes arbitrary coordinates $\mathbf{x}$ as input and predicts $\Phi$'s output $\mathbf{y} \in \mathrm{R}^{n_y}$. In contrast, mesh-based models, such as convolutional neural networks (CNNs), take an image $\mathbf{Y} = [\mathbf{y}_1, \cdots, \mathbf{y}_{n_t}]^\mathrm{T}$ as output, which contains fixed coordinates defined by a template mesh $\mathbf{X} =$ $[\mathbf{x}_1, \cdots, \mathbf{x}_{n_t}]^\mathrm{T}$, where $n_t$ is the number of data points in the template.

The key difference between mesh-agnostic models and mesh-based models concerns the data that they process, as demonstrated in Fig. 1. For example, when predicting flows around airfoils, the flow field data usually come from computational fluid dynamics (CFD) simulations, which can use structured grids, unstructured grids, hybrid grids or Cartesian grids. The mesh size can be different in different cases, and the mesh is usually dense near walls but coarse in the far field. The different mesh topologies and sizes are shown in the first row of Fig. 1. Generally, for classic mesh-based models such as vanilla CNNs and Autoencoders, the flow field data need to be reconstructed into a fixed and structured format, especially when the data are obtained from different sources. A very simplified example of data reconstruction is shown in Fig. 1, the orange lines are a uniform Cartesian template for CNNs, the flow field data on the C-grid is interpolated to the fixed template for training and utilization. It's usually difficult for structured templates to achieve high resolution in the boundary layer and shock wave region, especially for three-dimensional flows around complex geometries.

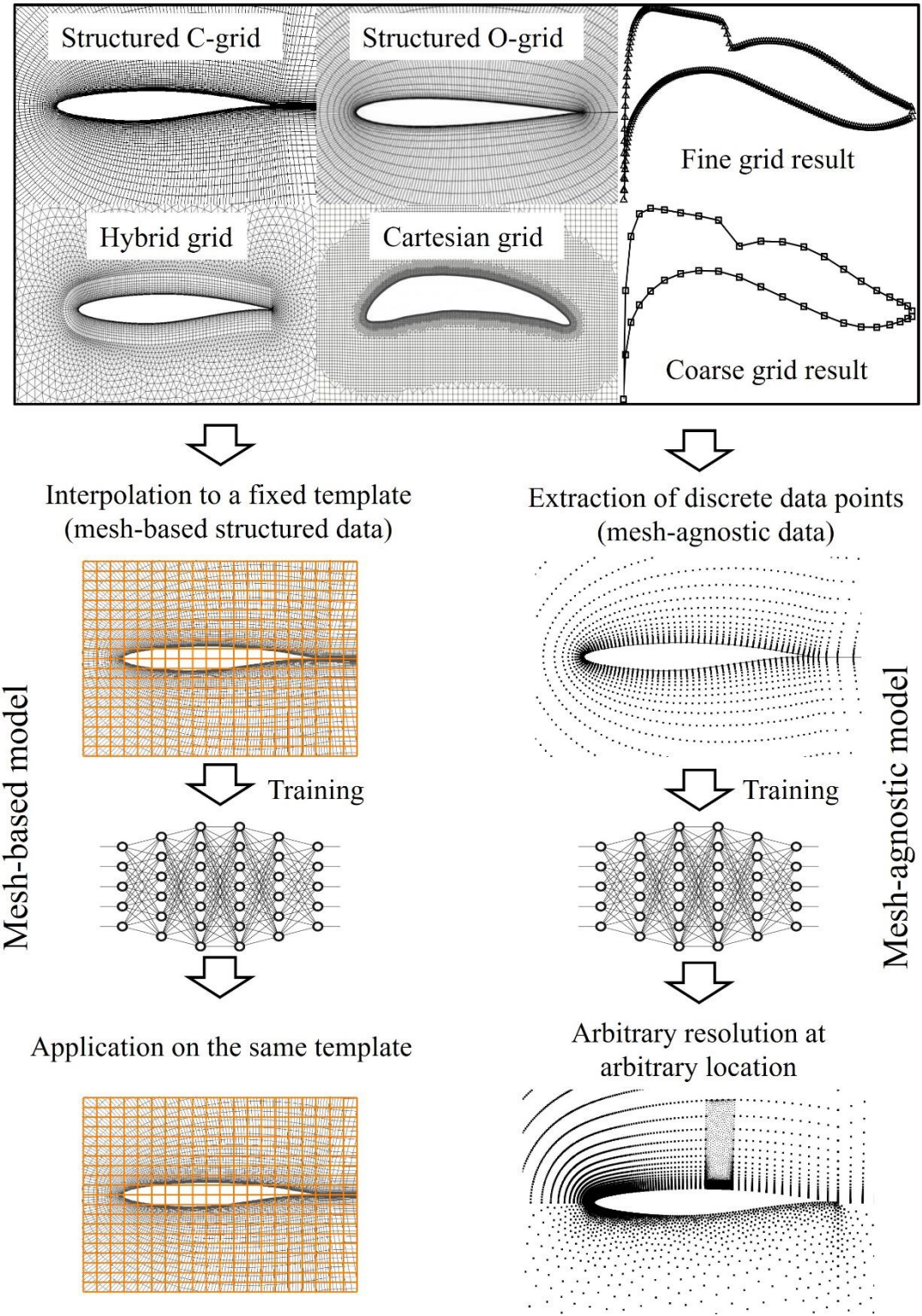

**Fig. 1 Difference between the data used by mesh-based and mesh-agnostic models**

In contrast, mesh-agnostic models process discrete data points in the flow field, so they are not dependent on the given data structure. Each data point in a flow field is a sample for the model. Therefore, the number and distribution of the data points are arbitrary, as shown in Fig. 1. Therefore, mesh-agnostic models have several advantages for tackling the aforementioned challenges: 1) the model complexity does not scale with the data complexity, so these models are more suitable for three-dimensional fields and unstructured data; 2) data from different meshes and sources can be used together, more data will help improve model generalization ability; 3) the PDE loss can be easily constructed because the partial derivatives can be addressed by automatic differentiation. Therefore, mesh-agnostic models avoid the discretization and meshing processes that are employed in conventional numerical methods or mesh-based models. So, it's easier to embed physical equations in the model.

Recently, there have been several machine learning algorithms developed that can directly use unstructured mesh samples, such as spherical CNNs [14], graph CNNs [15], and PointNet [16]. They have shown great potential for utilizing unstructured data and deserve further investigation. The key difference between these algorithms and the implicit decoder is that they use all data points in a flow field as one sample, but the implicit decoder uses each individual data point as a sample. There are pros and cons for these different approaches, the best option for different scenarios also needs further study.

Mesh-agnostic models have been extensively studied in recent years. For example, PINNs have been studied to model incompressible flows, turbulent flows, supersonic flows, geofluid flows, etc. [13,17]. The mesh-agnostic neural PDE solver (MAgNet) combines INR and graph neural networks to solve PDEs and achieves super resolution [18]. These models have been developed to solve PDEs. They need to be trained on a case-by-case basis and require training times that are comparable to those of classic solvers, e.g., CFD solvers [13].

However, fast prediction and design tools are desired in aerodynamic shape design cases. Therefore, it is necessary to combine mesh-agnostic models with traditional data-driven machine learning. Some studies have employed INR in data-driven architectures. For example, deep operator networks (DeepONets) [19] attach a hypernetwork [23] to INR, achieving a general architecture that can learn nonlinear operators. This method theoretically proves a universal approximation theorem of operators; it also shows that DeepONets achieve smaller generalization errors than fully connected neural networks (FNNs) in the fast prediction of one-dimensional dynamic system solutions. Neural implicit flow (NIF) [24] has a similar structure to DeepONets. It can conduct mesh-agnostic dimensionality reduction on spatial-temporal data in three-dimensional isotropic turbulence flows and can also predict unsteady two-dimensional cylinder flows. These data-driven mesh-agnostic models have demonstrated promising performance for simple flows, but their performance in terms of predicting flows around different geometric shapes is still unknown.

This paper proposes a data-driven mesh-agnostic model for the fast prediction and inverse design of supercritical airfoils. It uses INR to represent spatial bases in the flow field, thereby also enabling the use of the PDE loss for future physics-driven constraints. A hypernetwork is used to map the relationships between the model parameters (e.g., the free stream condition and airfoil geometry) and flow field bases. The proposed model can be utilized as a decoder in any conventional data-driven model while retaining the benefits of mesh-agnostic models. Therefore, it is called an implicit decoder in this paper, denoted by ImD.

This paper is organized as follows. First, the sample preparation process for supercritical airfoils is introduced. Next, the architecture of the implicit decoder is introduced, and different activation functions are discussed to improve the model generalization ability. Then, the model is tested to predict various supercritical airfoil shapes and pressure distributions. It is further utilized in a physically interpretable variational autoencoder for the inverse design of supercritical airfoils.

## II. Sample preparation

### A. Geometry and simulation of supercritical airfoils

Supercritical airfoils are constructed by the class shape transformation (CST) method [25], for which the shape function is a ninth-order Bernstein polynomial. The upper or lower surface of an airfoil is defined by

$$y = C_{N_2}^{N_1}(x) \sum_{i=1}^{n} c_i K_i x^i (1-x)^{n-i}, \ K_i = \frac{n!}{i!(n-i)!},$$
$$C_{N_2}^{N_1}(x) = x^{N_1}(1-x)^{N_2}, \ N_1 = 0.5, \ N_2 = 1.0, \quad (2)$$

where $n = 10$ for the ninth-order Bernstein polynomial, $x \in [0,1]$. Therefore, there are 20 CST parameters $\{c_i\}_{i=1}^{20}$ to define the unit chord length airfoil geometry.

A C-grid is generated for airfoils with unit chord lengths, and an open-source solver CFL3D [26] is used for the CFD simulations. $\Delta y+$ of the first grid layer is always set to be less than one. 301 grid points are distributed on the airfoil surface. The MUSCL scheme, Roe's scheme, the lower-upper symmetric Gauss-Seidel method, and the $k - \omega$ shear stress transport (SST) model are used in Reynolds average Navier–Stokes (RANS) simulations. The CFL number is 2.0 for 8000 steps. The CFD setup and validation processes are the same as those in [7].

The wall Mach number ($M_w$) distribution on the airfoil surface is extracted from the CFD results and used for training. It is the Mach number calculated based on an isentropic relationship with the pressure coefficient ($C_p$) on the airfoil surface and the free-stream Mach number $M_\infty$ [27]. The relationship is shown in Eq. (3), i.e.,

$$C_p = \frac{2}{\gamma M_\infty^2}\left[\left(\frac{2+(\gamma-1)M_\infty^2}{2+(\gamma-1)M_w^2}\right)^{\frac{\gamma}{\gamma-1}} - 1\right], \quad (3)$$

where the ratio of specific heats of air $\gamma = 1.4$. The wall Mach number values are generally more consistent than pressure coefficients with respect to physical features, e.g., shock waves. Because the value of critical $C_p$, where the local Mach number equals one, is influenced by the free stream Mach number. So, the wall Mach number reduces the difficulty of machine learning to learn the influence of $M_\infty$, and is therefore used in this paper.

Two physical features, i.e., the shock wave location $X_1$ and the wall Mach number in front of the shock wave $M_{w,1}$, are selected for inverse design purposes. The two physical features are shown in Fig. 2. The shock wave is recognized by an abrupt $M_w$ decrease on the upper surface, where $M_{w,1}$ exceeds one. The shock wave is first roughly located by the largest $-dM_w/dX$ location, $\tilde{X}$. Then, $M_{w,1}$ is defined as the $M_w$ at location $X_1$ in front of $\tilde{X}$, where $dM_w/dX|_{X_1} = -1$ or $d^2M_w/dX^2$ reaches a local minimum.

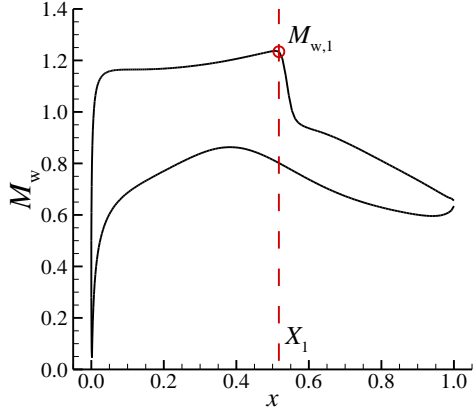

**Fig. 2 Definition of shock wave features (RAE2822 airfoil) [7]**

**B. Supercritical airfoil database**

There have been several databases for conventional airfoils and supercritical airfoils available online, but the number of supercritical airfoil samples are limited [28]. To improve the generalization ability of models, a supercritical airfoil database was built in [28] to cover the free-stream condition, airfoil geometry, and shock wave features that may occur in transonic civil transport aircraft. In this paper, the supercritical airfoil database is enlarged by adding the classic supercritical airfoils in Fig. 3 and another 100 conventional airfoils to the database. Fig. 4 shows some of the added conventional airfoils, these airfoils are selected from the UIUC low-speed airfoil database [28]. The enlarged database contains many airfoil samples with smaller trailing edge slope angle $\alpha_{TE}$, which increases the data diversity.

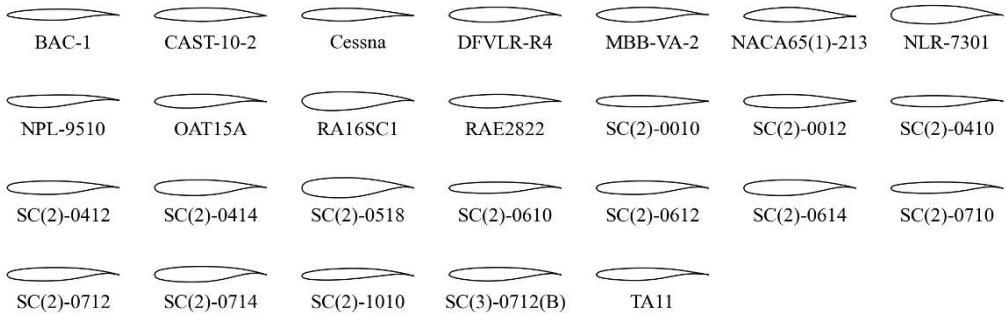

**Fig. 3 Classic supercritical airfoils [28]**

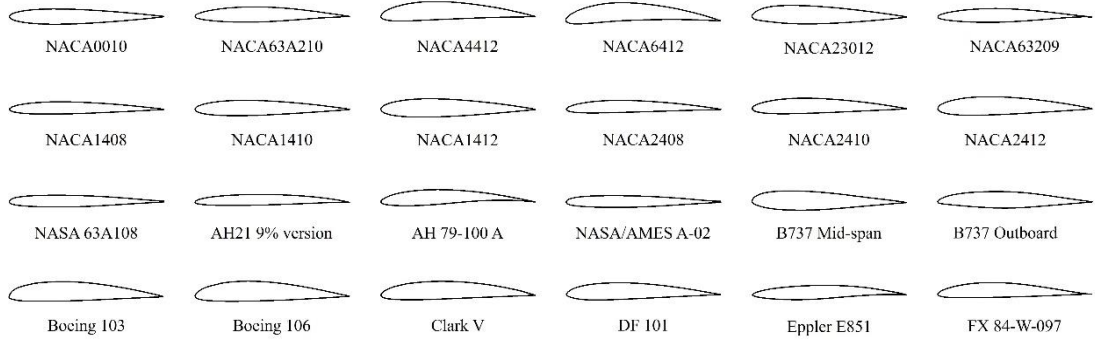

Fig. 4 Examples of the additional conventional airfoils

The final database contains 2,500 airfoils, some typical airfoils are shown in Fig. 5. Each row shows airfoils with different maximum relative thickness ($t_{max}$), the airfoil geometry ($y$) and wall Mach number distribution ($M_w$) are shown together, the corresponding free stream conditions are labeled on the top.

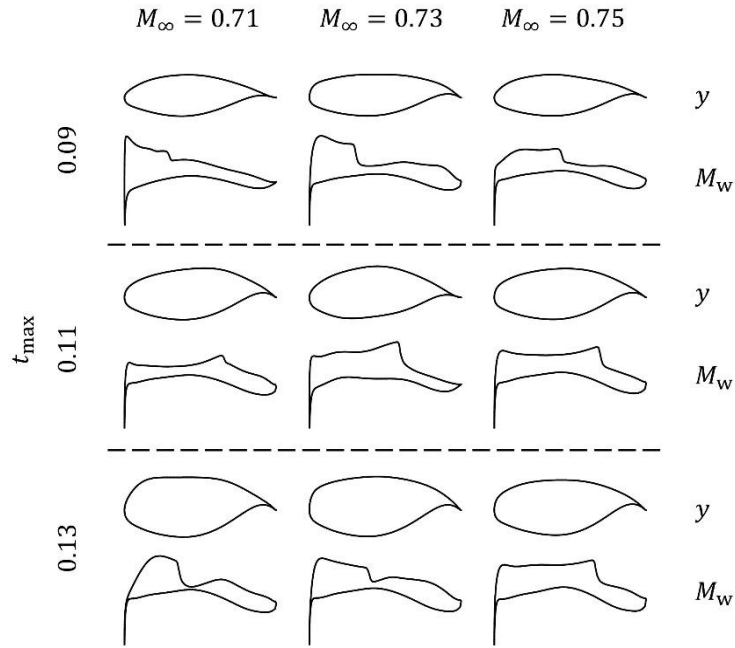

Fig. 5 Typical airfoils in the supercritical airfoil database

The sample density distributions are plotted in Fig. 6. Fig. 6(a)-(c) show the diversity of the free-stream conditions of airfoils, i.e., $M_\infty$, the $AoA$ (degree), and the lift coefficient $C_L$. Fig. 6(d)-(f) show the diversity of airfoil geometries, i.e., the maximum relative thickness $t_{max}$, leading edge radius $r_{LE}$, and trailing edge slope angle $\alpha_{TE}$ (degree). Fig. 6(g)-(i) show the diversity of the airfoil flow field, i.e., the shock wave location $X_1$, wall Mach number in front of the shock wave $M_{w,1}$, and moment coefficient $C_m$.

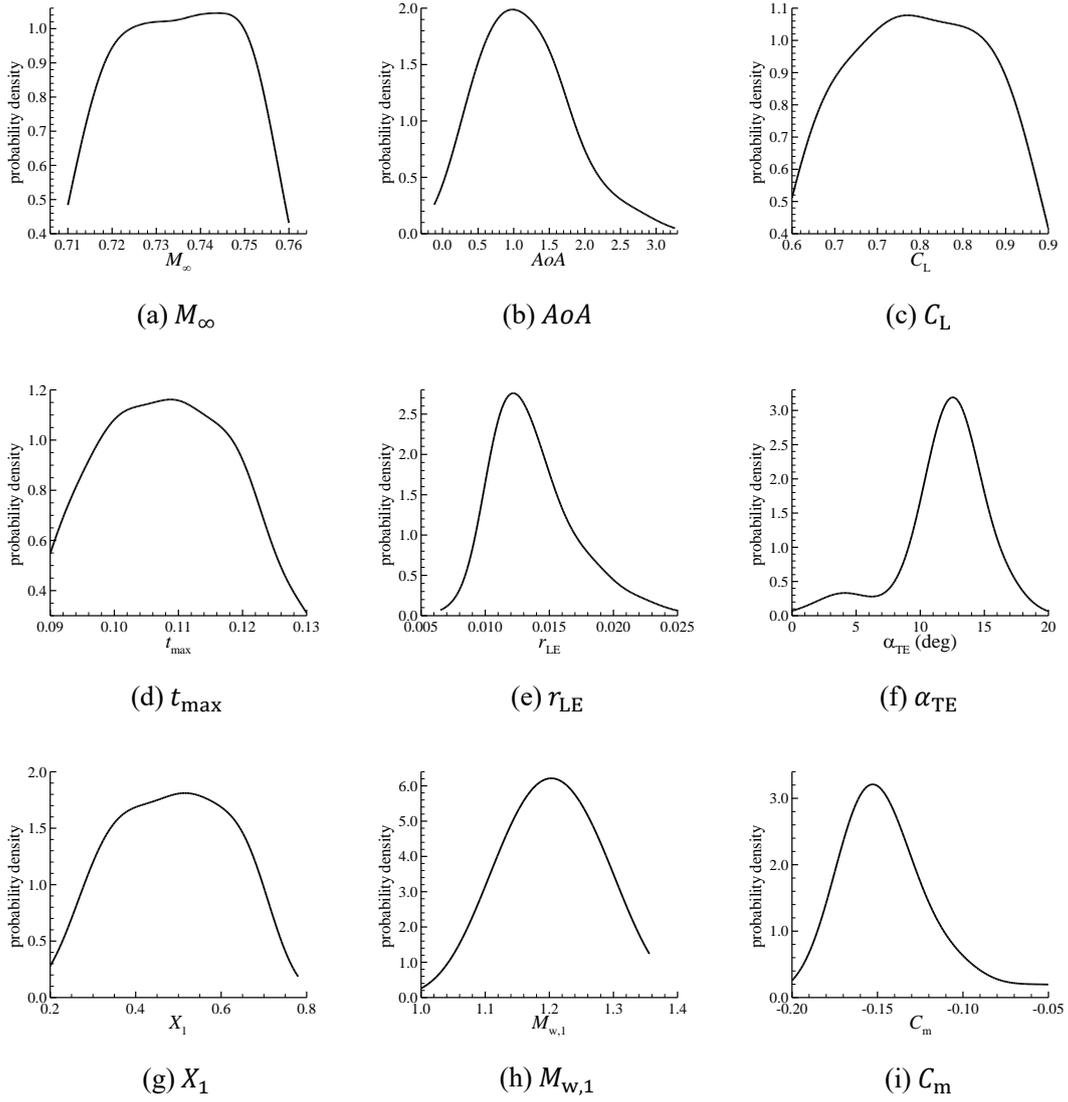

**Fig. 6 Sample distributions in the parametric spaces**

## C. Mesh-based and mesh-agnostic samples

As shown in Fig. 1, mesh-based and mesh-agnostic models have different data requirements. Mesh-based models process data that have the same size and structure; sometimes, the data also must be structured arrays. For example, classic data-driven models such as autoencoders (AEs) and CNNs are usually trained on mesh-based data. This brings great challenges for learning the spatial data of the flow fields around complex geometric shapes. In these scenarios, mesh-agnostic methods have greater advantages. This difference is caused by the fact that an entire data field is a sample for mesh-based models, but each data point is a sample for mesh-agnostic models.

Fig. 7 shows the difference between mesh-based and mesh-agnostic samples, using the airfoil wall Mach number distribution as an example. Fig. 7(a) shows two examples of reconstructed mesh-based airfoil samples. The mesh-based samples have the same number of data points, the data points also share the same order and the same *x*-coordinate distribution defined by a fixed template mesh **X**. The locations of points on the template mesh are plotted as dashed vertical lines. In contrast, mesh-agnostic airfoil samples may have different numbers of data points, and the data points can have different orders and different *x*-coordinate distributions, as shown in Fig. 7(b). For example, the data points can have larger density values near a shock wave to capture the associated discontinuity.

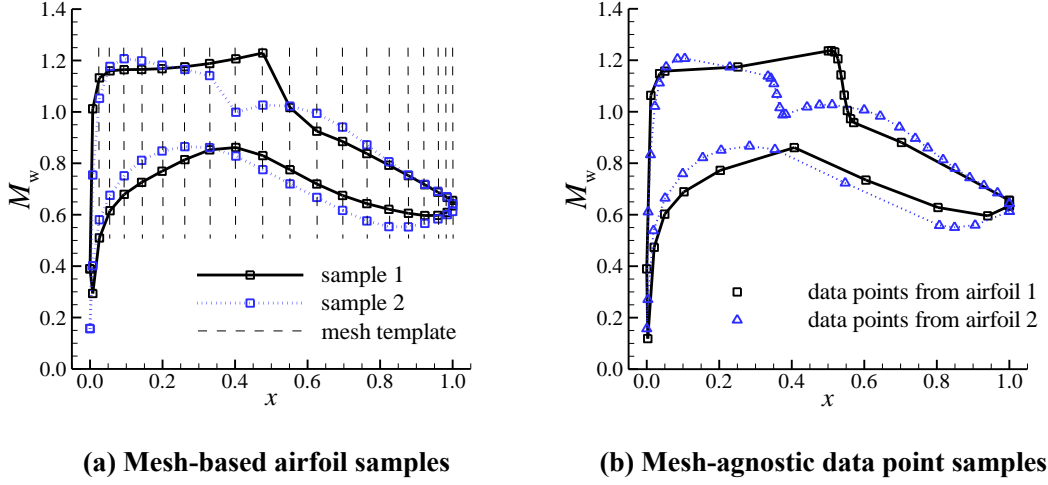

(a) Mesh-based airfoil samples  (b) Mesh-agnostic data point samples

Fig. 7 Mesh-based and mesh-agnostic airfoil samples

In this paper, samples for model training and testing consist of airfoil geometries $(x, y)$ and wall Mach number $(M_w)$ distributions. There are 201 data points distributed on the upper and lower surfaces of each airfoil sample, respectively. So, each airfoil contains 401 data points. The values of $y$ coordinate and $M_w$ are interpolated from the airfoil and CFD results, providing a given *x*-coordinate distribution. The *x*-coordinate distribution for mesh-based samples is the same, defined by

$$x = \frac{\left(\frac{1-\cos \pi a}{2}\right)^\beta - \left(\frac{1-\cos \pi a_0}{2}\right)^\beta}{\left(\frac{1-\cos \pi a_1}{2}\right)^\beta - \left(\frac{1-\cos \pi a_0}{2}\right)^\beta}, \ a = a_0(1-r) + a_1 r, \ r \in [0,1], \tag{4}$$

where $a_0 = 0.0079$, $a_1 = 0.96$, $\beta = 1.0$, and $r = (i-1)/n_{\text{point}}$ $(i = 1, \cdots, n_{\text{point}})$. $r$ is the ratio of the point index $i$ to the total number of points $(n_{\text{point}})$ in the *x* direction. Then, these airfoil samples are used for the training of mesh-based models, and the testing for both the mesh-based and mesh-agnostic models. On the other hand, mesh-agnostic models use individual data point as a sample. In this paper, 51 data points are randomly selected from the 201 data points in the upper and lower surfaces, respectively, both including the leading edge and trailing edge. Additionally, the data points within 5% of shock wave location $X_1$ on the upper surface are also included as the training samples for mesh-agnostic models. Therefore, the number and *x*-coordinate distribution of the data points extracted from each airfoil sample are not fixed, approximately 101 to 150 points are extracted from each airfoil.

## III. Theoretical methods

### A. Implicit decoder

The architecture of the proposed interpretable implicit decoder is shown in Fig. 8. It consists of two subnetworks: 1) *ShapeNet*, which represents spatial bases, and 2) *HyperNet*, which predicts the weight of each spatial base. The outputs of ShapeNet and HyperNet are multiplied in the *embedding layer*. The implicit decoder has a similar architecture to that of DeepONets [19]; the difference is that the trunk network of DeepONets is divided into ShapeNet and an embedding layer for better interpretability.

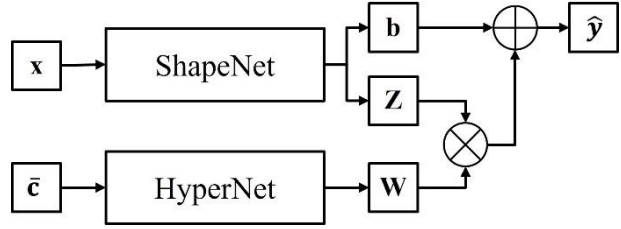

**Fig. 8 Architecture of the implicit decoder**

ShapeNet is an INR that takes coordinates $\mathbf{x}$ as input and predicts spatial bases $\mathbf{Z} \in \mathrm{R}^{n_l \times n_y}$ and the output bias $\mathbf{b} \in \mathrm{R}^{n_y}$. HyperNet is a hypernetwork that takes the coefficients $\bar{\mathbf{c}} \in \mathrm{R}^{n_{\bar{c}}}$ as inputs and predicts the weights $\mathbf{W} \in \mathrm{R}^{n_l \times n_y}$, where $n_l$ is the number of latent variables. It's worth noting that the number of spatial bases ($n_b$) equals the number of latent variables, i.e., $n_b = n_l$. Therefore, the number of spatial bases for the rest of this paper is denoted by $n_l$. Then, the output $\hat{\mathbf{y}}$ is calculated by Eq. (5):

$$\hat{y}_j = \sum_{k=1}^{n_l} W_{k,j} Z_{k,j} + b_j, \quad j = 1, \cdots, n_y, \tag{5}$$

where $j, k$ represent the indices of the components of vectors $\hat{\mathbf{y}}$ and $\mathbf{b}$, and matrices $\mathbf{W}$ and $\mathbf{Z}$. Eq. (5) shows that the output $\hat{\mathbf{y}}$ is a linear combination of the spatial bases $\mathbf{Z}$. This linear operator is usually called an embedding layer of ShapeNet and HyperNet.

The implicit decoder can be viewed as a function of $\mathbf{x}$ conditioned on $\bar{\mathbf{c}}$, which is denoted by

$$\hat{\mathbf{y}} = \Phi(\mathbf{x}|\bar{\mathbf{c}}), \tag{6}$$

where $\hat{\mathbf{y}}$ is the prediction of $\mathbf{y}$. Then, the data-driven loss function of the implicit decoder is

$$\text{loss}_{\text{data}} = \sum_{i=1}^{N_d} [\Phi(\mathbf{x}_i|\bar{\mathbf{c}}_i) - \mathbf{y}_i]^2, \tag{7}$$

where each data point provides a triplet $(\mathbf{x}_i, \bar{\mathbf{c}}_i, \mathbf{y}_i)$, and $N_d$ is the total number of data points.

It is worth noting that multiple data points are available in one airfoil sample defined by $\bar{\mathbf{c}}$. Denote $\{(\mathbf{x}_i, \bar{\mathbf{c}}_i, \mathbf{y}_i)\}_m$ as the data points extracted from the *m*-th sample, where $i = 1, \cdots, n_{d,m}$, $m = 1, \cdots, N_s$, and $N_s$ is the number of airfoil samples. Therefore, $N_d = \sum_{m=1}^{N_s} n_{d,m}$. Then, the 2,500 airfoils sampled in Section II.B provide more than 250,000 data point samples, since approximately 101 to 150 points are extracted from each airfoil. In this paper, $\mathbf{x}$ is the *x* coordinate, $\mathbf{y}$ is a tuple $(y, M_w)$ consisting of the $y$ coordinate and wall Mach number $M_w$, and $\bar{\mathbf{c}}$ may consist of airfoil geometric parameters (i.e., CST parameters), free stream conditions (i.e., $M_\infty$, $AoA$) or flow feature parameters (e.g., $M_{w,1}$, $X_1$). Both $y$ and $M_w$ are used as the model output to demonstrate that the implicit decoder can predict multiple outputs at the same time, and it is also useful to generate both the airfoil geometry and wall Mach number distribution during inverse design.

**B. Implicit decoder with interpretable modes**

As discussed in Section III.A, ShapeNet predicts the spatial bases, and HyperNet predicts the weights. Essentially, the implicit decoder first performs nonlinear dimensionality reduction and then makes predictions based on the bases. Nonlinear dimensionality reduction is often powerful but prone to overfitting, and the bases are usually difficult to interpret. Therefore, PCA constraints are applied to ShapeNet to improve the interpretability and generalization ability.

PCA is a popular linear dimensionality reduction method that projects data to a lower-dimensional subspace. It is often used for increasing interpretability and improving machine learning performance, by creating new uncorrelated variables that successively maximize variance [20]. PCA is a mesh-based algorithm, therefore, consider a template mesh $\mathbf{X} = [\mathbf{x}_1, \cdots, \mathbf{x}_{n_t}]^T \in \mathrm{R}^{n_t \times n_x}$ that consists of $n_t$ data points, the data field to be predicted $\mathbf{Y} = [\mathbf{y}_1, \cdots, \mathbf{y}_{n_t}]^T \in \mathrm{R}^{n_t \times n_y}$. When training implicit decoders on the template mesh, the same number of data points are extract from all airfoil samples based on the same template mesh. It means the implicit decoder is used in a mesh-based approach. Then, ShapeNet's outputs are $\bar{\bar{\mathbf{Z}}} \in \mathrm{R}^{n_t \times n_l \times n_y}$ and $\mathbf{B} \in \mathrm{R}^{n_t \times n_y}$, HyperNet's output is $\mathbf{W} \in \mathrm{R}^{n_l \times n_y}$. The double overbar on the uppercase variable indicates that it is a three-dimensional matrix. The output of the implicit decoder is

$$\hat{Y}_{i,j} = \sum_{k=1}^{n_l} W_{k,j} \bar{\bar{Z}}_{i,k,j} + B_{i,j}, \quad i = 1, \cdots, n_t, \ j = 1, \cdots, n_y, \tag{8}$$

where $i, j, k$ are indices.

It is worth noting that PCA has the same expression in Eq. (8). Denote the principal components of PCA by $\bar{\bar{\mathbf{P}}} = [\mathbf{P}_1, \cdots, \mathbf{P}_{n_p}] \in \mathrm{R}^{n_t \times n_p \times n_y}$ and the mean field by $\bar{\mathbf{Y}} \in \mathrm{R}^{n_t \times n_y}$, respectively. The overbar on the variable indicates that it is the mean value. Then, $n_p$ is the dimension of the linear subspace, which equals $n_l$. Therefore, by minimizing the PCA loss described by Eq. (9), i.e.,

$$\mathrm{loss}_{\mathrm{pca}} = \mathrm{MSE}(\bar{\bar{\mathbf{Z}}}, \bar{\bar{\mathbf{P}}}) + \mathrm{MSE}(\mathbf{B}, \bar{\mathbf{Y}}), \tag{9}$$

ShapeNet learns a mesh-agnostic orthogonal linear representation of the field $\mathbf{Y}$. MSE stands for the mean squared error.

Then, minimizing $\mathrm{loss}_{\mathrm{data}}$ together with $\mathrm{loss}_{\mathrm{pca}}$ makes ShapeNet capture the orthogonal bases on the template mesh, as well as other high-frequency structures in other spaces. In other words, the loss function of the implicit decoder with interpretable modes becomes

$$\text{loss} = \text{loss}_{\text{data}} + \text{loss}_{\text{pca}}, \qquad (10)$$

where $\text{loss}_{\text{data}}$ is the data-driven loss function of implicit decoders (Eq. (7)). When the template mesh is fine enough relative to the training data points, ShapeNet provides a sufficiently good set of bases for the problem. Then, HyperNet becomes a surrogate model for the coefficients acquired from PCA. But even when the template mesh is relatively coarse, the PCA constraint can still make the spatial bases orthogonal on the template mesh. Then, because the learned orthogonal spatial bases are usually easier to interpret than non-orthogonal bases [2, 20], using PCA constraints can make the model more interpretable. Furthermore, many adaptations of PCA have been proposed to improve physical interpretation of the reduced dimensions [20,21], which can be used in implicit decoder the same way as the classic PCA and further improve the interpretability.

**C. Mesh-agnostic variational autoencoders**

Variational autoencoders (VAEs) have an encoder-decoder architecture [22]. The encoder maps data to latent variables; then, the decoder reconstructs the data. To avoid the nonregularized latent spaces of autoencoders, VAEs force the latent distribution to be a standard high-dimensional normal distribution $N(\mathbf{I}_{n_l}, \mathbf{0})$. This provides the generative capability to the entire latent space, and $n_l$ is the dimension of the latent space.

VAEs are mesh-based models. The decoder of VAE can be easily replaced by the mesh-agnostic implicit decoder, but it is difficult to build a mesh-agnostic encoder. On the other hand, the encoder only needs to discern the differences between samples; therefore, a coarse snapshot of the sample should be sufficient for predicting the latent variables $\mathbf{v}$.

Each airfoil sample may have different numbers of data points available, i.e., their $n_d$ values are different. The data points may be evaluated at different coordinates from other samples, i.e., the $\mathbf{x}_i$s are different. Manually define a fixed template mesh $\mathbf{X} = [\mathbf{x}_1, \cdots, \mathbf{x}_{n_t}]^\text{T} \in \text{R}^{n_t \times n_x}$ that consists of $n_t$ data points, where $n_t \ll n_d$ for most samples. In this paper, the coarse template mesh has $n_t = 51$ data points, and $n_d$ is usually larger than 100. Extract a snapshot $\mathbf{Y}$ from all samples at $\mathbf{X}$ by interpolation. Then, each sample consists of its original data points $\{\mathbf{y}_i\}$ and a snapshot $\mathbf{Y}$. Notably, the template mesh needs to be carefully selected to capture the differences between samples.

The architecture of the mesh-agnostic VAE is shown in Fig. 9. The main part marked by black lines is used to train the VAE, so the template mesh $\mathbf{X}$ is fixed. The blue lines highlight the data flow for improving the implicit decoder. The loss function is described by Eq. (11).

$$\text{loss} = \text{loss}_{\text{re}} + \text{loss}_{\text{kld}} + \text{loss}_{\text{ma}}. \qquad (11)$$

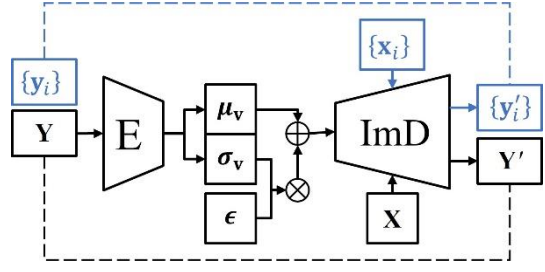

**Fig. 9 Architecture of the mesh-agnostic VAE**

VAE assumes that the data $\mathbf{Y}$ are generated by a random process $p(\mathbf{Y})$, involving unobserved continuous random latent variables $\mathbf{v}$. The process consists of two steps: 1) $\mathbf{v}$ is generated from a prior centered isotropic multivariate Gaussian $p(\mathbf{v}) = N(\mathbf{I_v}, \mathbf{0_v})$; 2) $\mathbf{Y}$ is generated from a conditional distribution $p_\theta(\mathbf{Y}|\mathbf{v})$. VAE lets the $p_\theta(\mathbf{Y}|\mathbf{v})$ be a multivariate Gaussian, whose distribution parameters are computed from $\mathbf{v}$. Then, the true posterior is $p_\theta(\mathbf{v}|\mathbf{Y}) = p_\theta(\mathbf{Y}|\mathbf{v})p(\mathbf{v})/p(\mathbf{Y})$. Since the true posterior is intractable, VAE lets the variational approximate posterior be a multivariate Gaussian with a diagonal covariance structure. It uses a reparameterization trick for the posterior, which is $\mathbf{v}_m \sim \boldsymbol{\mu}_{\mathbf{v},m} + \boldsymbol{\sigma}_{\mathbf{v},m} \odot \boldsymbol{\epsilon}$, $\boldsymbol{\epsilon} \sim N(\mathbf{0}, \mathbf{I})$, for the $m$-th sample ($m = 1, \cdots, N_s$). It means that the reparametrized latent variable $\tilde{\mathbf{v}}_m$ is sampled from the Gaussian distribution $N(\boldsymbol{\mu}_{\mathbf{v},m}, \boldsymbol{\sigma}_{\mathbf{v},m})$, where $(\mathbf{c}_m, \boldsymbol{\mu}_{\mathbf{v},m}, \boldsymbol{\sigma}_{\mathbf{v},m}) = \text{Encoder}(\mathbf{Y}_m)$. The tilted variable indicates that it is a reparametrized variable. The 're' in $\text{loss}_{\text{re}}$ stands for reparameterization, which maximizes the log-likelihood of the observed data $\mathbf{Y}_m$. The $\text{loss}_{\text{re}}$ is expressed by

$$\text{loss}_{\text{re}} = \sum_{m=1}^{N_s} \left\| \tilde{\mathbf{Y}}_m - \mathbf{Y}_m \right\|^2 / N_s, \quad \tilde{\mathbf{Y}}_m = \Phi(\mathbf{X}|\mathbf{v}_m). \tag{12}$$

The $\text{loss}_{\text{kld}}$ is the Kullback–Leibler divergence loss, minimizing the divergence of the approximate posterior from the exact posterior. The $\text{loss}_{\text{kld}}$ is expressed by

$$\text{loss}_{\text{kld}} = \sum_{m=1}^{N_s} \text{KL}[q(\mathbf{v}_m|\mathbf{Y}_m) \| p(\mathbf{v}_m)] / N_s, \tag{13}$$

where the Kullback–Leibler divergence (KLD) term is

$$\text{KL}[q(\mathbf{v}|\mathbf{Y}) \| p(\mathbf{v})] = \frac{1}{2} \sum_{j=1}^{n_v} \left[ 1 + \log \sigma_{\mathbf{v},(j)}^2 - \mu_{\mathbf{v},(j)}^2 - \sigma_{\mathbf{v},(j)}^2 \right]. \tag{14}$$

The index $j$ in Eq. (14) is the $j^{\text{th}}$ component of $\boldsymbol{\mu}_\mathbf{v}$ and $\boldsymbol{\sigma}_\mathbf{v}$. Then, the KLD term forces the latent distribution to be a standard high-dimensional normal distribution $N(\mathbf{I_v}, \mathbf{0_v})$. In other words, $\text{loss}_{\text{re}}$ and $\text{loss}_{\text{kld}}$ are the original VAE losses that apply to the snapshot $\mathbf{Y}$ [22].

For improving the implicit decoder, the $\text{loss}_{\text{ma}}$ is '$\text{loss}_{\text{re}}$' applied to $\{\mathbf{y}_i\}$, i.e.,

$$\text{loss}_{\text{ma}} = \sum_{m=1}^{N_s} \sum_{i=1}^{n_{d,m}} \left\| \tilde{\mathbf{y}}_{i,m} - \mathbf{y}_{i,m} \right\|^2 / n_{d,m} N_s, \tag{15}$$

where $\tilde{\mathbf{y}}_{i,m} = \Phi(\mathbf{x}_i|\mathbf{v}_m)$. The subscript 'ma' stands for mesh-agnostic data.

**D. Physically interpretable variational autoencoders**

To learn physically interpretable features with VAEs, the latent variables are divided into two parts, i.e., $(\mathbf{c}, \mathbf{v})$. The physical codes $\mathbf{c}$ are used to capture specific physical features $\mathbf{c}_0$, and the data features $\mathbf{v}$ are trained to represent the rest of the information in $\mathbf{Y}$. A physically interpretable variational autoencoder (PIVAE) [7] was designed to accurately reconstruct $\mathbf{Y}$, while $\mathbf{c}$ and $\mathbf{v}$ are independent of each other. Then, it can be used to generate new samples with specified physical features. The distribution of physical codes $p(\mathbf{c})$ is defined by the database, but the distribution of data features is a Gaussian distribution, i.e., $p(\mathbf{v}) = N(\mathbf{0}, \mathbf{I}_{n_v})$, which is the same as the classic VAE. Therefore, the decoder generates different possible $\mathbf{Y}$s that has the specified physical codes $\mathbf{c}$, by sampling $\mathbf{v}$.

The PIVAE architecture is shown in Fig. 10.

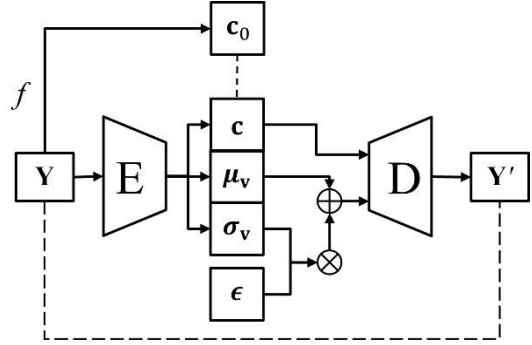

Fig. 10 Architecture of the PIVAE [7]

The loss function is

$$\text{loss} = \text{loss}_{\text{code}} + \text{loss}_{\text{mean}} + r(\text{loss}_{\text{id}} + \text{loss}_{\text{re}} + \text{loss}_{\text{kld}}). \tag{16}$$

The loss term of physical codes $\text{loss}_{\text{code}}$ encourages PIVAE to capture the physical codes, which is expressed by

$$\text{loss}_{\text{code}} = \sum_{m=1}^{N_s} \left\| \mathbf{c}_m - \mathbf{c}_{0,m} \right\|^2 / N_s. \tag{17}$$

The $\mathbf{c}_m$ in Eq. (17) is obtained by $(\mathbf{c}_m, \boldsymbol{\mu}_{\mathbf{v},m}, \boldsymbol{\sigma}_{\mathbf{v},m}) = \text{Encoder}(\mathbf{Y}_m)$, where $\mathbf{Y}_m$ is the ground truth snapshot of the $m$-th sample, $m$ is the sample index, and $N_s$ is the number of airfoil samples. In other words, the $\text{loss}_{\text{code}}$ describes the difference between the ground truth physical codes $\mathbf{c}_0$ and the predicted value $\mathbf{c}$ from the encoder.

The loss term of mean data reconstruction $\text{loss}_{\text{mean}}$ encourages PIVAE to accurately reconstruct the ground truth snapshot $\mathbf{Y}$ with the predicted physical code $\mathbf{c}$ and the predicted mean data feature $\boldsymbol{\mu}_{\mathbf{v}}$, where the mean snapshot $\overline{\mathbf{Y}} = \text{Decoder}(\mathbf{c}, \boldsymbol{\mu}_{\mathbf{v}})$. The $\text{loss}_{\text{mean}}$ is expressed by

$$\text{loss}_{\text{mean}} = \sum_{m=1}^{N_s} \| \overline{\mathbf{Y}}_m - \mathbf{Y}_m \|^2 / N_s. \tag{18}$$

In other words, the $\text{loss}_{\text{mean}}$ describes whether the decoder can accurately reconstruct the snapshot based on the output of encoder, **c** and $\boldsymbol{\mu_v}$. The $\text{loss}_\text{re}$ and $\text{loss}_\text{kld}$ in Eq. (16) are the loss functions of the original VAEs, i.e., Eq. (12) and (13). The ratio $r$ slowly increases from a small number, e.g., 0.0001, to a constant value, e.g., 0.1, so that the algorithm converges better. The details of PIVAE can be found in [7].

A loss function for the independence between $(\mathbf{c}, \mathbf{v})$ is added to PIVAE, and it is expressed as follows:

$$\text{loss}_\text{id} = \sum_{m=1}^{N_s} \left\| \tilde{\mathbf{c}}_m - \mathbf{c}_{0,m} \right\|^2 / N_s, \quad (19)$$

where $\mathbf{c}_{0,m}$ is the ground truth physical codes, $\tilde{\mathbf{c}}_m$ is the physical codes of $\widetilde{\mathbf{Y}}_m$ extracted by the encoder, i.e., $(\tilde{\mathbf{c}}_m, *, *) = \text{Encoder}(\widetilde{\mathbf{Y}}_m)$. $\widetilde{\mathbf{Y}}_m$ is the predicted snapshot based on $\mathbf{c}_{0,m}$ and the reparametrized data features $\tilde{\mathbf{v}}_m \sim N(\boldsymbol{\mu}_{\mathbf{v},m}, \boldsymbol{\sigma}_{\mathbf{v},m})$, i.e., $\widetilde{\mathbf{Y}}_m = \text{Decoder}(\mathbf{c}_{0,m}, \tilde{\mathbf{v}}_m)$. The $\text{loss}_\text{id}$ encourages $E_{\mathbf{v},\mathbf{v}' \sim p(\mathbf{v})} \|\mathbf{c}(\mathbf{v}) - \mathbf{c}(\mathbf{v}')\| \to 0$. Then, $\forall \mathbf{v}, \mathbf{v}'$, $p(\mathbf{c}|\mathbf{v}) = p(\mathbf{c}|\mathbf{v}')$. Therefore, $\forall \mathbf{v}$, $p(\mathbf{c}) = \int_{\mathbf{v}'} p(\mathbf{v}')p(\mathbf{c}|\mathbf{v}')d\mathbf{v}' = p(\mathbf{c}|\mathbf{v}) \int_{\mathbf{v}'} p(\mathbf{v}')d\mathbf{v}' = p(\mathbf{c}|\mathbf{v})$, and **c** is independent of **v**. In other words, the $\text{loss}_\text{id}$ describes whether the predicted physical codes **c** will change when the data features **v** are randomly sampled. Therefore, PIVAE with $\text{loss}_\text{id}$ no longer requires $p(\mathbf{c}) = 1$ in the training sample set, in contrast to the algorithm in [7].

## IV. Prediction of supercritical airfoils

A common application of implicit decoders is the construction of predictive models for forward problems. In this section, different activation functions are compared for ShapeNet. Next, the implicit decoder is used to predict supercritical airfoils. Then, the benefit of introducing the PCA constraint for the generalization ability of the model is discussed.

### A. Comparison between different activation functions for ShapeNet

Many studies have been conducted on the activation functions of INRs, and the common activation functions, e.g., ReLU, tanh, and sigmoid functions, have been proven inefficient for learning high-frequency functions [29,30]. Some studies have stated that multilayer perceptrons (MLPs) with common activation functions may be problematic when the given flow field has high-frequency structures or discontinuities [24,31]. A $\omega_0$-scaled sine activation function (called SIREN [10], where $\omega_0$ is often 30) is usually suggested in these cases, e.g., turbulent flows and multiscale flows. However, few high-frequency structures are contained in Reynolds-averaged flow field around airfoils, and the sizes and activation functions of ShapeNet are still worth discussing.

As introduced in Section III.A, ShapeNet represents the spatial bases of the flow field of interest. It can learn a mesh-agnostic complete orthogonal basis set by minimizing $\text{loss}_\text{pca}$ when the template mesh is fine and the latent dimension $n_l$ is sufficiently large. Therefore, the PCA modes can be used as training data to test the performance of ShapeNet. The geometry and wall Mach number distribution of airfoil samples are reconstructed on an equally-spaced mesh with 51 grid points. The lower surface of the airfoil is concatenated with the upper surface in its reverse order, and the $x$ coordinates of the data points on the lower surface are set to their negative values. Therefore, the $x$ coordinates of the reconstructed samples range from $[-1,1]$, the trailing edge of the lower surface is at -1, the leading edge is at 0, the trailing edge of the upper surface is at 1, and the template mesh size is $n_t$=101. The $y$ coordinates and $M_w$ values of the airfoil samples are used for PCA and ShapeNet training. Note that the output $\{y, M_w\}$ is normalized to $[0,1]$ for training and testing ML models. ShapeNet's latent dimension $n_l$ is set to 20, which equals the number of principle components. The percentage of variance explained by the PCA components is shown in Fig. 11; the blue line is the percentage of each component, and the black line is the summation of the first several components. The figure shows that 10 components represent over 90% of the information contained in the data, and 20 components cover over 99% of the information. The mean value $\bar{y}$ and first eight principal components $\mathbf{p}$ are shown in Fig. 12. The results also indicate that no highly efficient structures are observed.

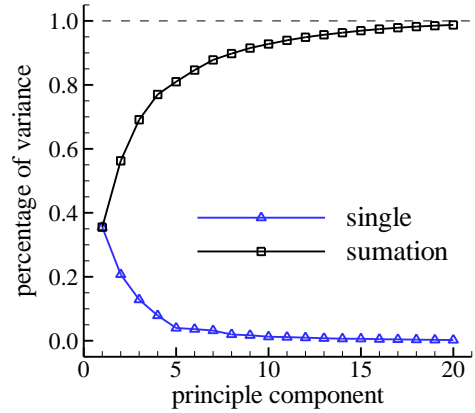

**Fig. 11 Percentage of variance explained by the PCA components ($n_t$=101)**

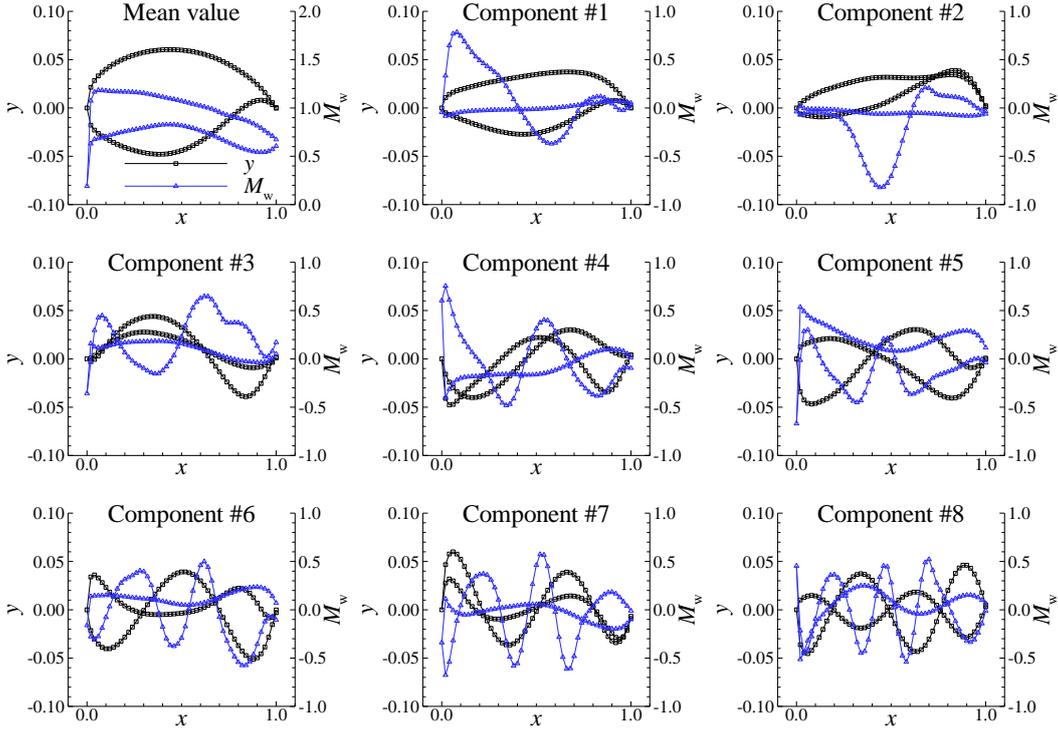

**Fig. 12 Mean values of the airfoil samples and PCA components**

Then, ShapeNet is trained on the PCA components by minimizing the loss in Eq. (9), and the model is updated by the Adam optimizer [32] for a maximum of 100,000 epochs. The learning rate starts from 0.001 for the ReLU and sigmoid activation functions and from 0.0001 for the sine activation function, and it gradually decreases to $1 \times 10^{-6}$ during training. It should be noted that MLPs with sine activation functions need a specific initialization strategy, which can be found in [10]. The MSE losses of ShapeNet variants with different sizes, depths, and activation functions are listed in Table 1. The first column shows the MLP depth $a$ and the number of neurons $b$ in each hidden layer, denoted by $a \times b$. The unsatisfying cases are marked by blue text.

**Table 1 MSE losses of ShapeNet with different PCA bases ($n_t$=101, $n_l$=20)**

| Hidden structure | Sine ($\omega_0$=3) | Sine ($\omega_0$=10) | Sine ($\omega_0$=30) | ReLU | Sigmoid |
|---|---|---|---|---|---|
| 1×32  | $4 \times 10^{-4}$ | $2 \times 10^{-5}$ | $4 \times 10^{-6}$ | $2 \times 10^{-4}$ | $3 \times 10^{-4}$ |
| 1×64  | $6 \times 10^{-5}$ | $6 \times 10^{-7}$ | $2 \times 10^{-7}$ | $2 \times 10^{-5}$ | $3 \times 10^{-5}$ |
| 1×128 | $2 \times 10^{-6}$ | $2 \times 10^{-7}$ | $2 \times 10^{-8}$ | $1 \times 10^{-5}$ | $1 \times 10^{-5}$ |
| 3×32  | $4 \times 10^{-6}$ | $8 \times 10^{-7}$ | $8 \times 10^{-7}$ | $4 \times 10^{-5}$ | $5 \times 10^{-6}$ |
| 3×64  | $4 \times 10^{-7}$ | $4 \times 10^{-8}$ | $5 \times 10^{-8}$ | $5 \times 10^{-6}$ | $7 \times 10^{-7}$ |
| 3×128 | $1 \times 10^{-7}$ | $2 \times 10^{-8}$ | $1 \times 10^{-9}$ | $3 \times 10^{-7}$ | $8 \times 10^{-8}$ |

This indicates that traditional activation functions usually require slightly larger MLPs to fit PCA bases than those of the sine activation function. Fig. 13 shows the mean values and principal components predicted by ShapeNets with different activation functions. These ShapeNets have the same hidden layer structure ($3 \times 64$) and similar MSE losses as well. The results show that both the sigmoid and sine ($\omega_0$=3) functions achieve smooth bases, but sine ($\omega_0$=30) overfits the data and results in a noisy prediction. Furthermore, the high-frequency noise is more severe for coarse template meshes. For example, when the template mesh size $n_t$ equals 21, the mean values predicted by ShapeNets are plotted in Fig. 14. The exact values are plotted as scatters. The hidden layer structure of the ShapeNets is $1 \times 64$, and the models are all well-trained. It is clear that sine ($\omega_0$=30) causes unwanted high-frequency noise. Therefore, traditional activation functions can be used for modeling flows around airfoils, and their predictions are usually relatively smooth. The sine activation function achieves good results with small $\omega_0$ values, but the risk of overfitting rapidly increases with larger $\omega_0$ values, especially when the template mesh size $n_t$ is small.

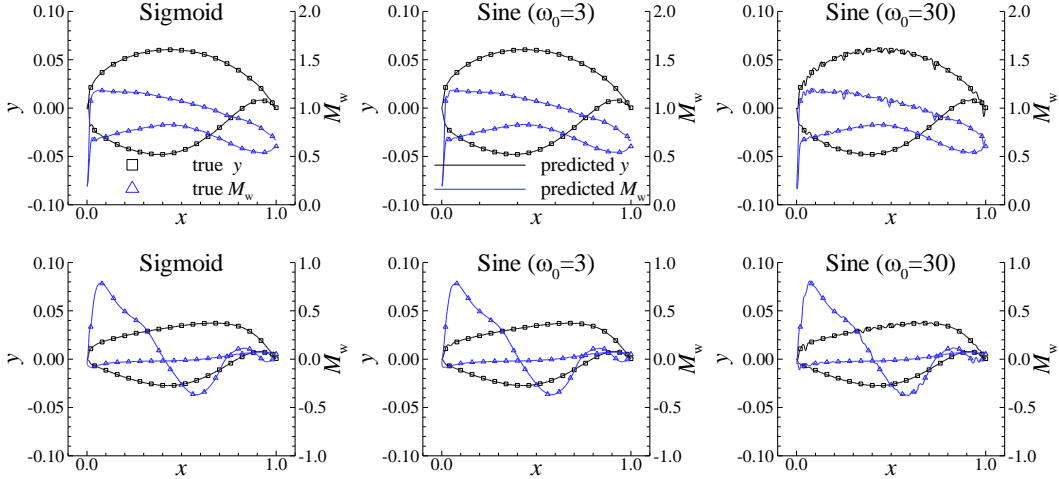

**Fig. 13 Spatial bases predicted by ShapeNets with different activation functions ($n_t$=101)**

**(One of every four true value scatters is plotted for clarity.)**

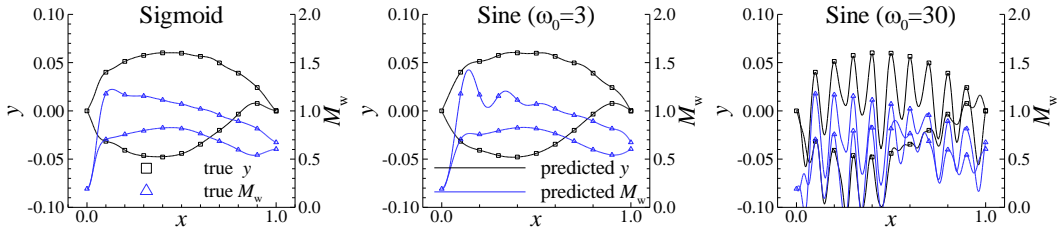

**Fig. 14 Predictions of ShapeNets on a coarse template with different activation functions ($n_t$=21)**

**B. Airfoil prediction based on implicit decoders**

The implicit decoder is employed to predict supercritical airfoils based on the free-stream condition $\{M_\infty, AoA\}$ and CST parameters $\{c_i\}_{i=1}^{20}$, i.e., $\bar{\mathbf{c}} = \{M_\infty, AoA, c_i\}, i = 1, \cdots, 20$. The model architecture is demonstrated in Fig. 15. ShapeNet consists of three hidden layers with 256 neurons, and its activation function is the sigmoid function. HyperNet consists of three hidden layers with 256 neurons, and its activation function is a ReLU. The input of ShapeNet is the $x$ coordinate of an arbitrary data point, and the final outputs are the $y$ coordinate and wall Mach number $M_w$ of the predicted supercritical airfoil at that location. Since the implicit decoder is a mesh-agnostic model, the number and $x$ coordinates of the data points in each airfoil sample are not fixed, which brings great flexibility to the training and prediction processes.

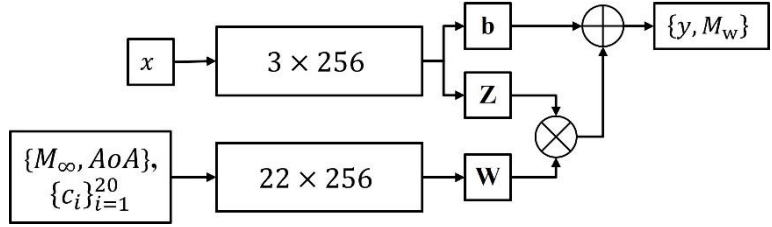

**Fig. 15 Architecture of the implicit decoder used for airfoil prediction**

The implicit decoder is trained by Adam for a maximum of 200,000 epochs, and all samples are fed into the model in one batch at each epoch. Ninety percent of the airfoil database is randomly selected as the training set, and the other airfoils compose the testing set. The learning rate starts from 0.001 and gradually decreases to $1 \times 10^{-6}$ during training. The loss function is described in Eq. (10). During model tuning, the size of ShapeNet is found to be larger than those needed in Section IV.A to achieve small prediction errors. The number of latent variables $n_l$ in the ShapeNet-predicted spatial bases $\mathbf{Z} \in \mathrm{R}^{n_l \times n_y}$ is 20, and the number of principle components $n_p$ is also 20 for $\mathrm{loss}_{\mathrm{pca}}$ in Eq. (9). The template mesh size for PCA is $n_t$=101. The loss history is plotted in Fig. 16.

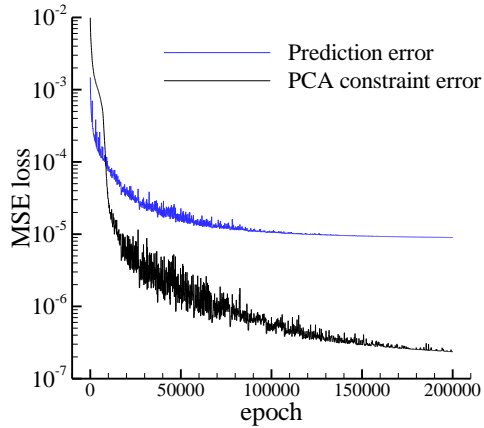

**Fig. 16 Loss history of the implicit decoder for airfoil prediction**

Fig. 17 shows the spatial bases predicted by ShapeNet. The scatter points are the true values derived from PCA, and the solid lines are the predicted values. The results show that no high-frequency noise is contained in the predicted bases. Fig. 18 shows the predicted supercritical airfoils for the training set (first row) and testing set (second row). The root mean square error (RMSE) of airfoil $\{y, M_w\}$ on the training set is 0.22%, and the RMSE induced on the testing set is 0.63%. The RMSE values are the average values of five individual training sessions.

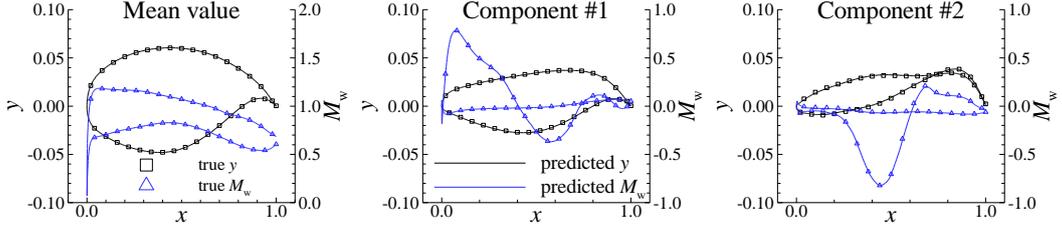

**Fig. 17 Spatial bases predicted by the implicit decoder with PCA constraints ($n_t$=101)**
**(One of every four true value scatters is plotted for clarity.)**

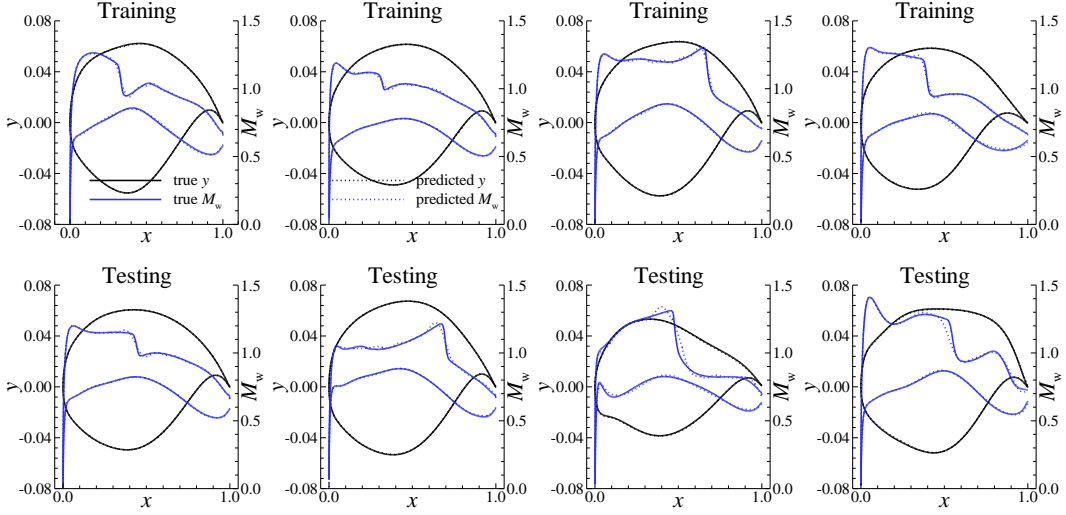

**Fig. 18 Supercritical airfoils predicted by the implicit decoder with PCA constraints**
**(Solid: true data; dashed: predicted data)**

## C. Influences of the latent dimension and the PCA constraint

In this section, the influence of the latent dimension when the PCA constraint is activated is discussed. As introduced in Section III.B, the loss function of the implicit decoder with the PCA constraint is described in Eq. (10), i.e., $\text{loss} = \text{loss}_{\text{data}} + \text{loss}_{\text{pca}}$. The difference between using orthogonal, non-orthogonal bases, or a combination of both are studied in this section. The ShapeNet and HyperNet structures are kept the same as those in Section IV.B.

When the PCA constraint is activated and $n_l = n_p$, the spatial bases of ShapeNet become the principal components by minimizing the PCA loss described by Eq. (9). As shown in Fig. 11, 20 principal components cover over 99% of the information, but ten components cover only 90% and five components cover only about 80%, which are not sufficient for accurately reconstructing all details in the sample set. Five cases are studied to reveal the influence of latent dimension ($n_l$) to the implicit decoder for airfoil prediction. The template mesh size for PCA ($n_t$) is 101, the number of data points for prediction ($n_d$) is 401, the number of principle components ($n_p$) equals $n_l$.

The RMSE losses of implicit decoders are shown in Fig. 19. Each RMSE value is the average value of five individual training sessions. The bule and red lines show the RMSE of data-driven loss function on the training and testing airfoil data points, i.e., $\text{loss}_{\text{data}}$ in Eq. (7), which is the prediction error of the airfoil geometry and wall Mach number distribution. The black line shows the RMSE of PCA constraint, i.e., $\text{loss}_{\text{pca}}$ in Eq. (9), which is the error between the ShapeNet's spatial bases and the principal components. The results show that the prediction accuracy is low when the latent dimension is small. This is the limit of linear dimensionality reduction methods, which often require a larger number of bases to reconstruct data than nonlinear methods.

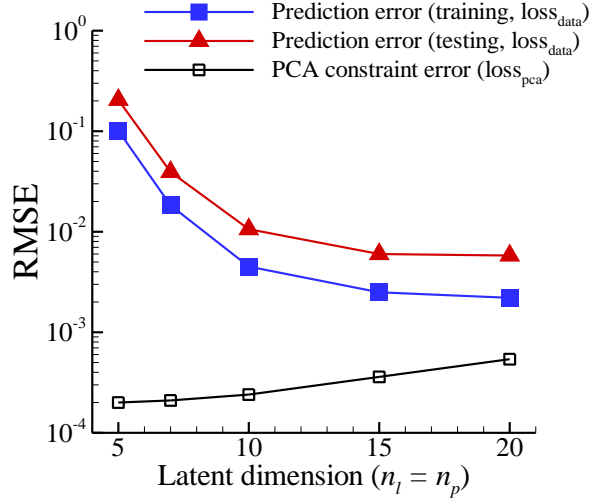

**Fig. 19 RMSE losses of implicit decoders with different latent dimensions**

On the other hand, nonlinear dimensionality reduction methods are more likely to suffer from overfitting. Table 2 lists detailed settings of three cases that have different combination of orthogonal and non- orthogonal bases. The template mesh size for PCA ($n_t$) is 101, the number of data points for prediction ($n_d$) is 401. Each RMSE value is the average value of five individual training sessions. Case 1 is the result of $n_l = n_p = 10$ in Fig. 19, case 2 frees two latent variables to learn nonlinear spatial bases, and case 3 uses all latent variables to learn nonlinear bases. Results show that the prediction error of case 2 is smaller than case 1, it achieves similar performance with the model using 20 linear bases in Fig. 19. Case 3 has a larger prediction error on the testing set than case 2. Fig. 20 shows the predicted supercritical airfoils for the testing set, each row belongs to one case. Although the RMSEs induced by cases 1 and 3 on the testing set seem small (approximately 1%), the details of the shock wave and lower surface are not accurately captured.

Therefore, combining the orthogonal PCA bases with non-orthogonal bases can achieve a balance between dimensionality reduction and generalization ability. The orthogonal bases capture the major proportion of information, improving model interpretability and avoiding severe overfitting in the unseen samples, while the non-orthogonal bases enable the model to capture the details of different samples.

**Table 2 RMSE losses of implicit decoders**

| Case | PCA constraint | Latent dimension ($n_l$) | Number of principle components ($n_p$) | Prediction RMSE (training set) | Prediction RMSE (testing set) | PCA RMSE |
|---|---|---|---|---|---|---|
| 1 | activated | 10 | 10 | 0.51% | 1.06% | 0.02% |

| 2 | activated | 10 | 8 | 0.32% | 0.60% | 0.04% |
| 3 | deactivated | 10 | -- | 0.31% | 0.95% | -- |

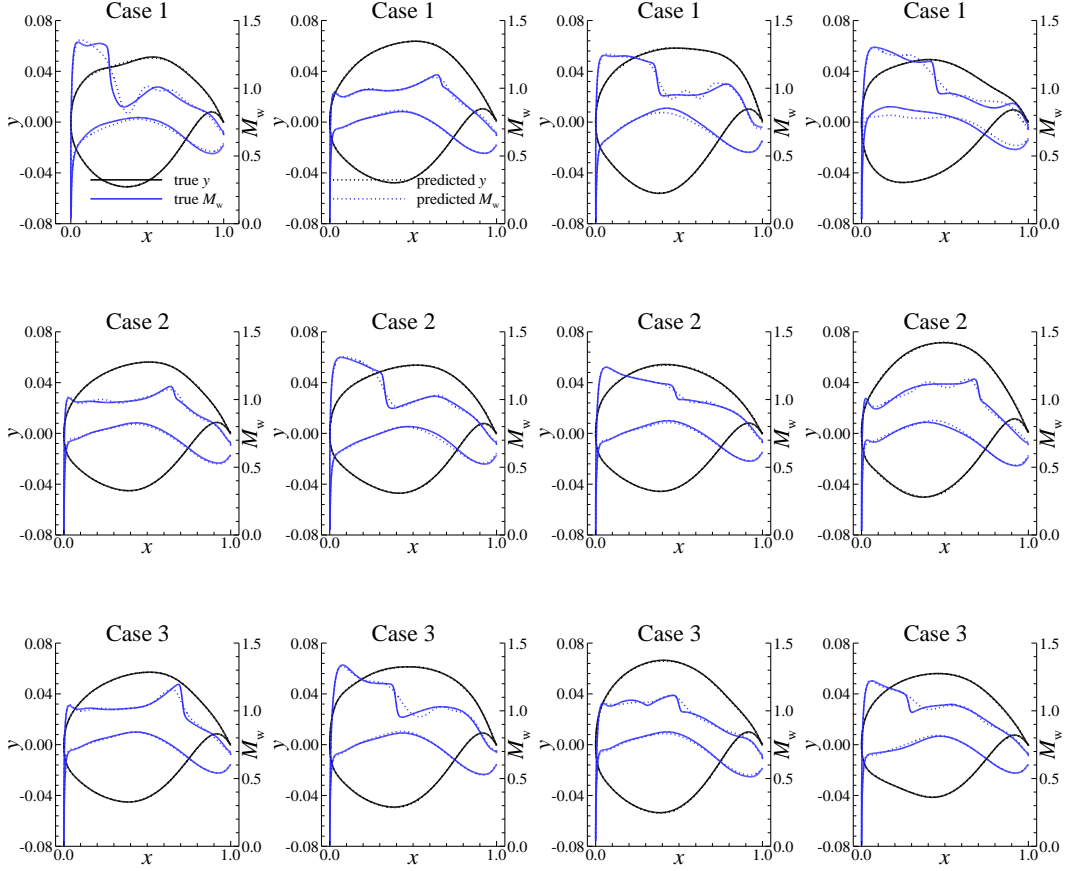

**Fig. 20 Generalization ability comparison among implicit decoders**
**(Solid: true data; dashed: predicted data; each row shows the results obtained by different model setups on the samples in the testing set)**

### D. Comparison with the mesh-based prediction model

The mesh-based model for airfoil geometry and wall Mach number distribution is easy to build, a vanilla fully connected neural network is used for prediction. The airfoil samples for training are introduced in Section II. C, which consists of 401 data points. The $y$ coordinate and $M_\text{w}$ are interpolated from the airfoil and CFD results at the fixed template mesh. Therefore, the model output is $\{(y, M_\text{w})_j\}_{j=1}^{401}$, where $j$ is the index of data points. The model input is the free stream condition and airfoil CST parameters $\{c_i\}_{i=1}^{20}$, where $i$ is the index of CST parameters. The model architecture is shown in Fig. 21, which contains three hidden layers, the hidden layers have 128, 512 and 1024 neurons, respectively. The airfoils in the training set and testing set are the same as the implicit decoder. The model is trained by Adam for a maximum of 5,000 epochs, the learning rate starts from $1 \times 10^{-2}$ and gradually decreases to $1 \times 10^{-4}$ during training. The loss function is the MSE between the airfoil sample and model output, the MSE loss is reduced to $2.8 \times 10^{-6}$ after training. The RMSE of airfoil $\{y, M_w\}$ on the training set is 0.17%, and the RMSE induced on the testing set is 0.43%. The RMSE values are the average values of five individual training sessions. The RMSE is similar to the results of well-trained mesh-agnostic model.

Fig. 22 shows the predicted supercritical airfoils by the mesh-based model in the testing set, the airfoils are the same four airfoils for Case 2 in Fig. 20, which represent the best performance of implicit decoders. The results indicate that the implicit decoders achieve similar results as the mesh-based models when they share the same training set, but it usually takes more epochs and smaller learning rate for training than mesh-based models.

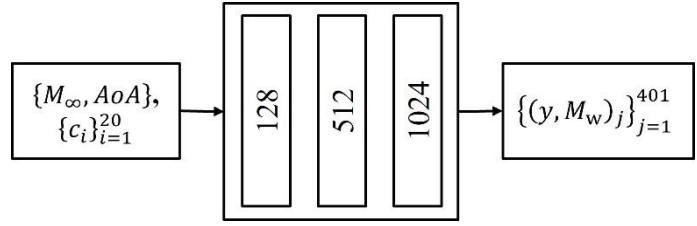

**Fig. 21 Architecture of the mesh-based neural network for airfoil prediction**

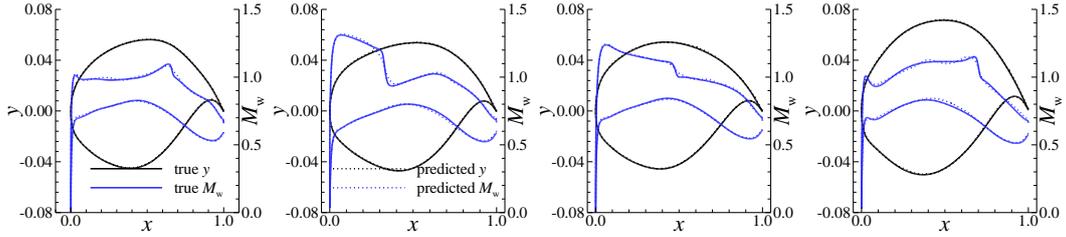

**Fig. 22 Performance of the mesh-based neural network on the testing airfoil samples**
**(Solid: true data; dashed: predicted data)**

## V. Inverse design of supercritical airfoils

In this section, the implicit decoder is used for the inverse design of airfoils. The implicit decoder takes the free stream condition $M_\infty, AoA$, design objectives, and the $x$ coordinate as input, and airfoil geometry ($y$ coordinate) and wall Mach number ($M_w$) as the output. The shock wave location $X_1$ is chosen as the design objective. The difference between implicit decodes in airfoil inverse design and airfoil prediction (Section IV) is that the input of model contains design objectives for inverse design rather than geometry parameters for prediction. The output of models is the same for both scenarios, because it helps demonstrate the performance of implicit decoders.

### A. Mesh-agnostic PIVAE for inverse design

The implicit decoder is used for inverse design, and the architecture of this mesh-agnostic PIVAE is shown in Fig. 23. Each airfoil sample consists of three parts, i.e., mesh-agnostic data $\{(y, M_w)_i\}_{i=1}^{n_{d,m}}$, a snapshot $\mathbf{Y} = \{(y, M_w)_i\}_{i=1}^{n_t}$ at the fixed coarse template mesh, and physical codes $\mathbf{c} = \{M_\infty, AoA, X_1\}$ for inverse design purposes. The mesh-based coarse snapshot and physical codes are used for the PIVAE training process, which is the same as the standard training process in [7]. The mesh-agnostic data are used to refine ShapeNet to capture airfoil details, and its data flow is highlighted by blue lines in Fig. 23. The encoder is an MLP with three hidden layers, and the activation function is a ReLU. ShapeNet consists of three hidden layers with 256 neurons, and its activation function is the sigmoid function. HyperNet consists of three hidden layers with 256 neurons, and its activation function is a ReLU. The implicit decoder is trained by Adam for a maximum of 200,000 epochs, and all samples are fed into the model in one batch at each epoch. Ninety percent of the airfoil database is randomly selected as the training set, and the other airfoils compose the testing set. The learning rate starts from $1 \times 10^{-4}$ and gradually decreases to $1 \times 10^{-6}$ during training.

The number of latent variables $n_l$ in the ShapeNet-predicted spatial bases $\mathbf{Z} \in \mathrm{R}^{n_l \times n_y}$ is ten, and the number of principle components $n_p$ is eight. The template mesh size for the snapshot is $n_t$=41. The dimension of the physical codes ($n_c$) is three, and the dimension of the data features ($n_v$) is eight. The loss history is plotted in Fig. 24.

Fig. 25 shows the model predictions for the mesh-based coarse snapshot, the mesh-agnostic data, and the actual CFD mesh points. The first row shows the results of RAE2822, the second row concerns OAT15A, and the reconstruction MSE loss induced on these two samples is approximately $10^{-6}$. The first column shows the coarse snapshot for the encoder-decoder training; therefore, the data are mesh-based data and consist of fewer data points. The number of data points in the snapshot is $n_t$=41. The second column shows the mesh-agnostic data used to refine ShapeNet, which can provide higher resolutions near the shock wave and leading edge. The number of data points is about 120, as introduced in Section II. C. The third column shows the model predictions produced for the actual CFD mesh points (301 data points). Therefore, mesh-agnostic models have great sample flexibility and thus a better ability to focus on important details.

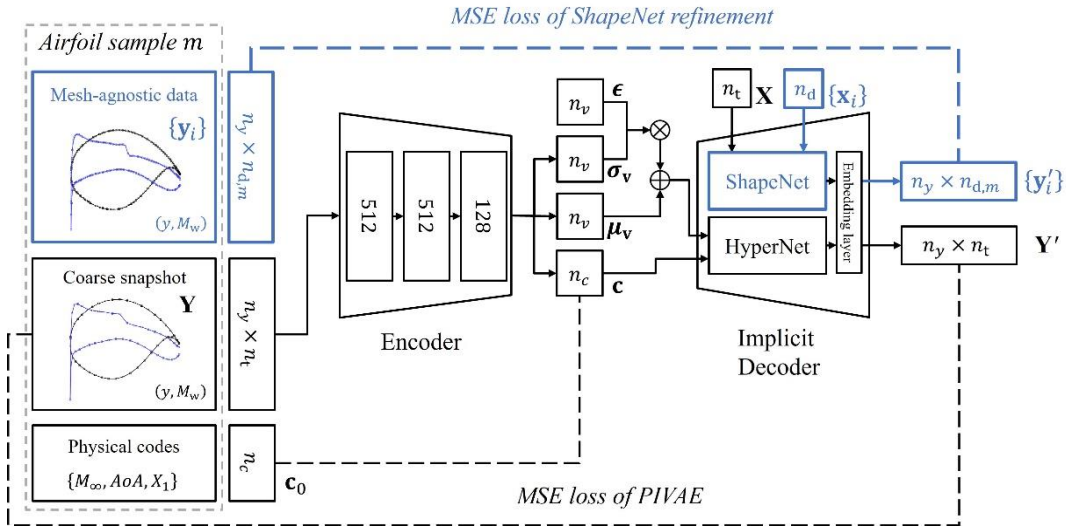

**Fig. 23 Architecture of the mesh-agnostic PIVAE for inverse airfoil design**

**(Black lines: data flow for the PIVAE training process with coarse snapshots; blue lines:**

data flow for refining ShapeNet with mesh-agnostic data; dashed lines: MSE loss functions)

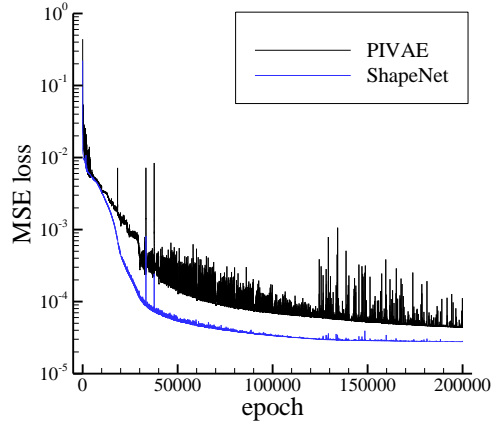

Fig. 24 Loss history of the mesh-agnostic PIVAE for inverse airfoil design

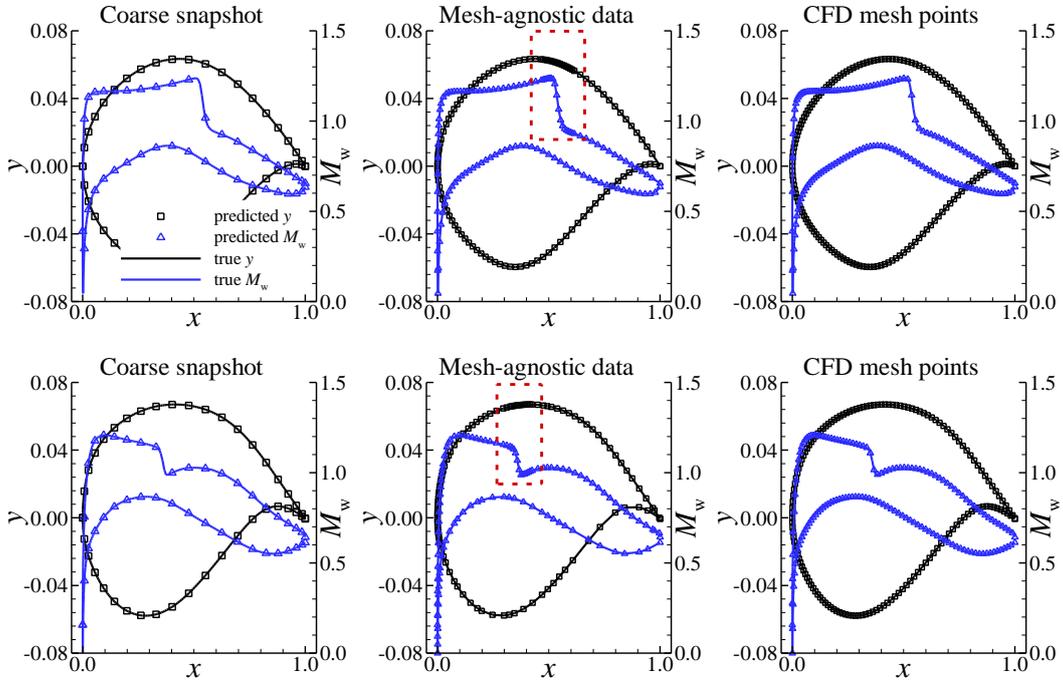

Fig. 25 Mesh-agnostic data for model training and airfoil reconstruction

(First row: RAE2822; second row: OAT15A)

## B. CFD validation of the inverse design results

The trained model is used for the inverse design of supercritical airfoils with specific physical codes (**c**); $M_\infty$ equals 0.72 or 0.74; the $AoA$ equals 1.0 or 1.5; and the shock wave location $X_1$ equals 0.3, 0.5, or 0.7. The data features (**v**) are set to zero so that the model generates airfoil geometries that achieves the specified physical codes. The model predicts the corresponding airfoil geometries (black lines) and wall Mach number distributions (blue lines), which are plotted in Fig. 26. The airfoil geometries are then used for CFD validation, and the true wall Mach number distributions (orange lines) are also plotted in Fig. 26. The $y$ coordinates of airfoil geometry are multiplied by ten and added with 0.5 for better plotting. The actual airfoil geometries are plotted in

Fig. 27. The results show that the implicit decoder can be employed as the decoder in a VAE and achieve good inverse design accuracy.

Fig. 28 plots the CFD results obtained for inversely designed supercritical airfoils, and the physical codes are the same, i.e., $M_\infty = 0.74$, $AoA = 1.5$, $X_1 = 0.5$. The black solid lines are the results of $\mathbf{v} = 0$. The orange dashed lines are the other possible airfoils, i.e., the $\mathbf{v}$ is randomly sampled. The results show that the actual physical codes, i.e., the shock wave locations $X_1$, are not influenced by random values of the data features $\mathbf{v}$.

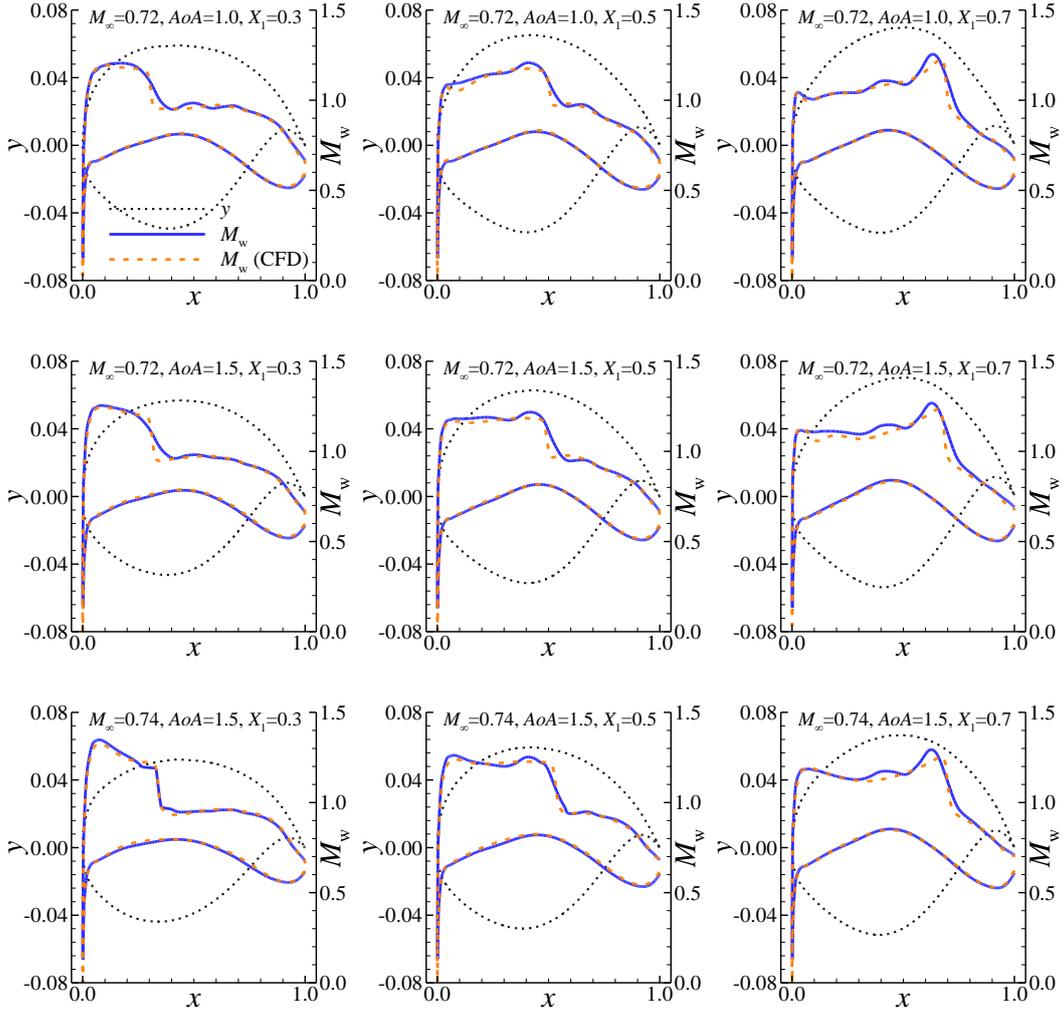

**Fig. 26 CFD validation of the inversely designed airfoils (v=0)**

**(Solid: generated geometries and wall Mach number distributions;**

**dashed: CFD validation results obtained on the generated geometries)**

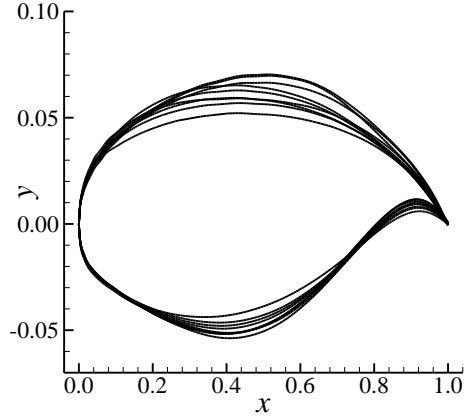

Fig. 27 Geometries of the generated airfoils (v=0)

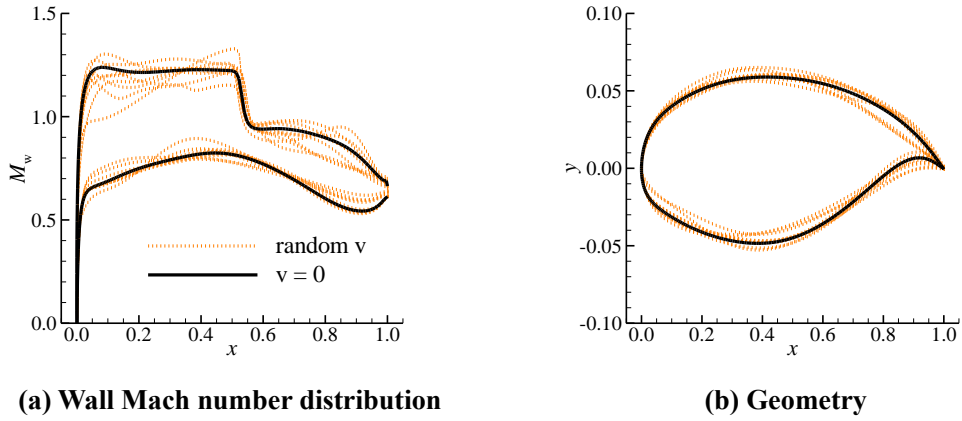

(a) Wall Mach number distribution  (b) Geometry

Fig. 28 CFD validation of the independence between  c  and  v

(Solid: $v = 0$; dashed: random  v)

## C. Comparison with the mesh-based PIVAE for inverse design

The mesh-agnostic PIVAE replaces the decoder in the original mesh-based PIVAE [7] by an implicit decoder. In this section, a mesh-based PIVAE is trained on the same training set for comparison. The airfoil samples for training are introduced in Section II.C, which consists of 401 data points. The $y$ coordinate and $M_w$ are interpolated from the airfoil and CFD results at the fixed template mesh. Therefore, the input of encoder and output of decoder is $\{(y, M_w)_j\}_{j=1}^{401}$, where $j$ is the index of data points. The model architecture is shown in Fig. 29, the encoder and decoder have three hidden layers. The airfoils in the training set are the same as the mesh-agnostic PIVAE. The learning rate of the Adam optimizer is 0.0002 for 10,000 epochs, and the mini-batch size is 64. The MSE loss of PIVAE is the same as the mesh-agnostic PIVAE, which is reduced to $8.1 \times 10^{-5}$ after training.

Both the mesh-agnostic and mesh-based PIVAE are employed to generate airfoils with different $X_1$s under different free stream conditions. The difference of $X_1$ between the specified value and the actual value extracted from the CFD results is used to measure the performance of inverse design models. The test airfoils are generated using the following method: 1) 21 values of the design physical code $X_1$ are evenly selected in the range of [0.20,0.75]; 2) two combinations of the remaining physical codes are randomly sampled, i.e., $\{M_\infty, C_L\}$; 3) for each combination of physical codes, one sample is generated with its data feature $\mathbf{v} = 0$, four samples are generated with randomly sampled $\mathbf{v}$. Therefore, 210 samples are generated and validated by CFD. The actual physical codes are then extracted from the CFD result and compared with the model input.

The results are shown in Fig. 30. There are 21 values evenly distributed in the $X_1$ axis, ten samples are included in the vertical line of each $X_1$. The solid scatters are the samples with $\mathbf{v} = 0$, the empty scatters are the samples with randomly sampled $\mathbf{v}$. Therefore, the vertical distance from each scatter to the dashed line represents the error of the inverse design sample, the deviation of the empty scatters in each vertical line indicates the disentanglement between the physical code $X_1$ and the data features. The results indicate that the mesh-agnostic model achieves similar performance as the mesh-based model when they share the same training set. Both models achieve good capability for the inverse design of airfoils for specific shock wave locations, the disentanglement between the physical codes and data features is also achieved.

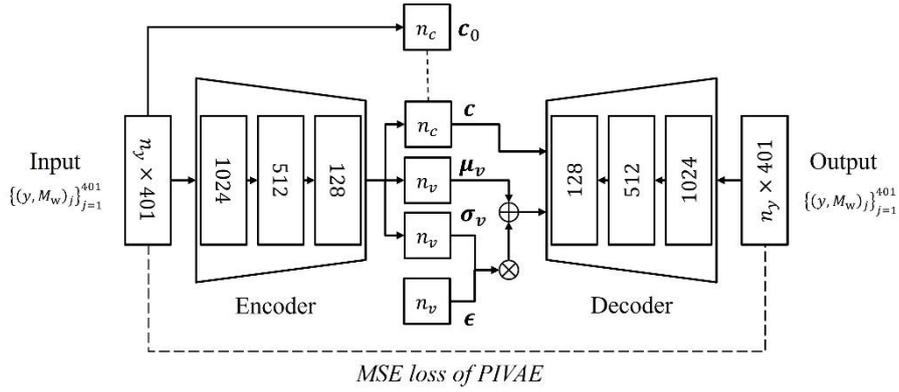

**Fig. 29 Architecture of the mesh-based PIVAE for inverse airfoil design**
**(Solid lines: data flow for the PIVAE training; dashed lines: MSE loss functions)**

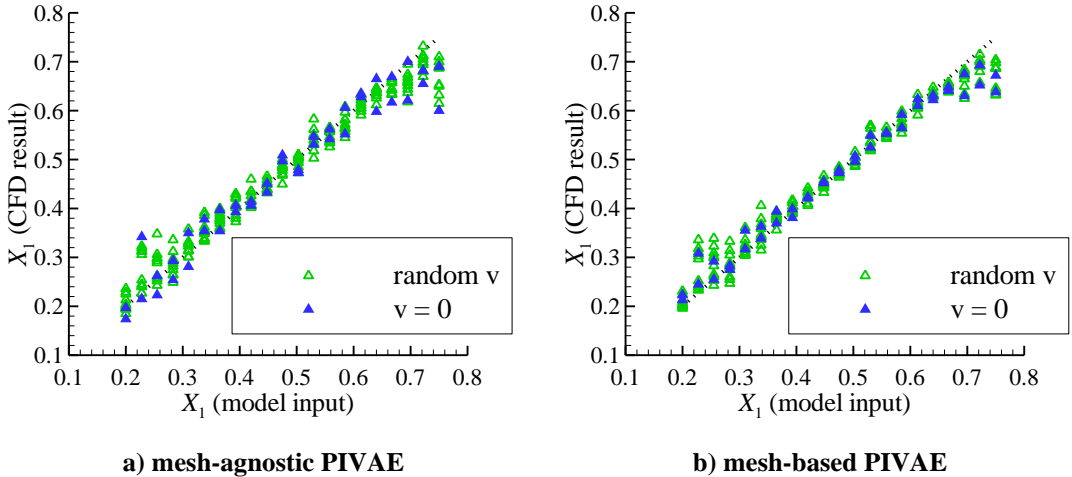

a) mesh-agnostic PIVAE         b) mesh-based PIVAE

**Fig. 30 Difference between the $X_1$ inputs and CFD results**

(Solid scatter: $v = 0$; empty scatter: randomly sampled $v$)

## VI. Conclusion

In recent years, mesh-agnostic models have shown advantages in terms of incorporating partial differential equations and great potential for processing unstructured three-dimensional spatial data. Therefore, they have been studied to solve PDEs and predict flows in/around simple geometries. This paper proposes a data-driven mesh-agnostic model called an implicit decoder (ImD) for the fast prediction of different supercritical airfoils. The influences of different activation functions and PCA constraints are discussed. Then, the implicit decoder is used to build a generative model for the inverse design of supercritical airfoils.

The following conclusions are reached.

1) The implicit decoder consists of two subnetworks, i.e., ShapeNet and HyperNet. ShapeNet is an implicit neural representation (INR) that predicts spatial bases and enables mesh-agnostic capabilities. The airfoil geometry or flow field is a linear combination of the spatial bases. HyperNet predicts the weight of each spatial base according to the input of the decoder. Then, the ImD can be used as a decoder in any classic model, such as an autoencoder.

2) Periodical activation functions (sine functions) are usually suggested for enabling INRs to capture high-frequency structures or discontinuities. Since the Reynolds-averaged flow fields around airfoils do not have many high-frequency structures, traditional activation functions, such as the ReLU and sigmoid function, can also be used in the ImD. Traditional activation functions achieve smoother results, but they need larger MLPs than the sine function. On the other hand, the sine function can achieve good results with small $\omega_0$ values, but the risks of overfitting and producing high-frequency noises rapidly increase with larger $\omega_0$ values.

3) ShapeNet predicts the spatial bases of flow fields, which are usually non-orthogonal bases because of the non-linear nature of neural networks. PCA is a popular linear dimensionality reduction method that extracts orthogonal spatial bases which describe most of the variation in the data. Therefore, ShapeNet can learn orthogonal spatial bases by adding a PCA constraint loss function, to improve the physical interpretability as well as the generalization ability of the ImD. Combining orthogonal bases and non-orthogonal bases can achieve good generalization with a relatively small number of latent variables.

4) The architecture of a mesh-agnostic PIVAE is proposed, which combines a mesh-agnostic ImD and a mesh-based encoder. The model is trained for the inverse design of supercritical airfoils with specific physical codes. Airfoil snapshots extracted from a fixed coarse template mesh are used as the encoder inputs, and the associated mesh-agnostic data are used to refine ShapeNet to capture flow field details. CFD results validate the accuracy of the obtained inverse design results, as well as the independence between the physical codes and data features.

5) When being trained on the same samples, mesh-agnostic models achieve similar

performance as mesh-based models, but they usually need more epochs and smaller learning rate than mesh-based models.

In summary, the proposed implicit decoder can be used as a mesh-agnostic decoder in any typical application of a conventional decoder. The activation function in ShapeNet should be determined based on the given data spatial structures, and the application of PCA constraints can help improve the generalization ability of the model.

**Acknowledgments**

This work was supported by the National Natural Science Foundation of China under Grant Nos. 92052203, 12202243, 92152301 and 91952302.